\documentclass[12pt]{article}

\usepackage{jheppub}
\usepackage{amsbsy,amsfonts,amsmath,amssymb}
\usepackage{cleveref}
\usepackage[latin1]{inputenc}

\crefname{equation}{}{}
\crefname{chapter}{Chapter}{Chapters}
\crefname{section}{Section}{Sections}
\crefname{subsection}{Subsection}{Subsections}
\crefname{subsubsection}{Subsubsection}{Subsubsections}
\crefname{figure}{Figure}{Figures}
\crefname{table}{Table}{Tables}
\crefname{appendix}{Appendix}{Appendices}

\renewcommand{\arraystretch}{1.5} 

\allowdisplaybreaks

\title{Second order transport coefficients of nonconformal relativistic fluids in various dimensions from Dp-brane}

\author[]{Chao Wu}

\affiliation[]{MTA Lend\"{u}let Holographic QFT Group, Wigner Research Centre for Physics, H-1525 Budapest 114, P.O.B. 49, Hungary}

\emailAdd{chaowu86@outlook.com}

\abstract{We derive all the dynamical second order transport coefficients for Dp-brane with $p$ from 1 to 6 within the framework of fluid/gravity correspondence in this paper. The D5 and D6-brane do not have dual relativistic fluids; D3-brane corresponds to 4-dimensional conformal relativistic fluid; D1, D2 and D4-brane separately correspond to nonconformal relativistic fluids of dimensions 2, 3 and 5. The Haack-Yarom relation only exists for Dp-branes with $p$ larger than 2 and is also satisfied by them. We also find that the Romatschke and Kleinert-Probst relations need to be generalized in order to be valid for relativistic fluids of dimensions other than 4.}


\arxivnumber{1807.08268}

\begin{document}

\maketitle

\section{Introduction}

Relativistic hydrodynamics is a widely used tool in theoretical physics, but its second-derivative behavior is not well understood. This behavior is mostly studied in two-formalisms that are both developed in 2000s. They are the Minkowskian prescription of AdS/CFT duality \cite{Son0205,Policastro0205,Policastro0210,Baier0712,Barnes1004,Arnold1105} and the fluid/gravity correspondence \cite{Bhattacharyya0712,Bhattacharyya0803,Bhattacharyya0806,Haack0806,Bhattacharyya0809,Erdmenger0809}. Much progress has been made in calculating transport coefficients in both of these formalisms, for various kinds of relativistic fluids. We are mainly focusing on the fluids without conserved vector current in this paper.

To review what is known, we begin by writing down the most general stress-energy tensor for an uncharged, nonconformal, relativistic fluid in $d$ dimensions. This is
\begin{align}\label{eq: relativistic fluid 2nd general stress tensor}
  T_{\mu\nu} =&\; \varepsilon u_\mu u_\nu + \mathfrak p P_{\mu\nu} - \left( \eta \cdot 2\sigma_{\mu\nu} + \zeta P_{\mu\nu} \nabla_\rho u^\rho \right) + \eta\tau_\pi \cdot 2 \left( \sideset{_\langle}{}{\mathop D} \sigma_{\mu\nu\rangle} + \frac{\sigma_{\mu\nu} \nabla_\rho u^\rho}{d-1} \right) \cr
  & + \eta \tau_\pi^* \cdot \frac{2 \sigma_{\mu\nu} \nabla_\rho u^\rho}{d-1} + \kappa \left( R_{\langle\mu\nu\rangle} - 2 u^\rho u^\sigma R_{\rho\langle\mu\nu\rangle\sigma}\right) + \kappa^* \cdot 2 u^\rho u^\sigma R_{\rho\langle\mu\nu\rangle\sigma} \cr
  & + \lambda_1 \cdot 4\sigma_{\langle\mu}^{~~\rho}\sigma_{\nu\rangle\rho} + \lambda_2 \cdot 2 \sigma_{\langle\mu}^{~~\rho}\Omega_{\nu\rangle\rho} + \lambda_3 \Omega_{\langle\mu}^{~~\rho}\Omega_{\nu\rangle\rho} + \lambda_4 \nabla_{\langle\mu} \ln s \nabla_{\nu\rangle} \ln s \cr
  & + P_{\mu\nu} \Big(\zeta\tau_\Pi D(\nabla_\rho u^\rho) + \xi_1 \cdot 4\sigma_{\rho\lambda}\sigma^{\rho\lambda} + \xi_2 (\nabla_\rho u^\rho)^2 + \xi_3 \Omega_{\rho\lambda} \Omega^{\rho\lambda}  \cr
  & + \xi_4 P^{\rho\lambda} \nabla_\rho \ln s \nabla_\lambda \ln s + \xi_5 R +  \xi_6 u^\rho u^\lambda R_{\rho\lambda} \Big),
\end{align}
where $\varepsilon$ is the energy density and $\mathfrak p$ is the pressure. The metric $g_{\mu\nu}$ gives the Ricci and Riemann tensor $R_{\mu\nu}$ and $R_{\mu\nu\rho\sigma}$ as usual. $P_{\mu\nu}=g_{\mu\nu}+u_\mu u_\nu$ is the spatial projection tensor, and $D = u^\mu \nabla_\mu$ is the temporal derivative, and the spatial-projected traceless symmetric tensor is defined as
\begin{align}
  A_{\langle\mu\nu\rangle} = P_\mu^\rho P_\nu^\lambda A_{(\rho\lambda)} - \frac1{d-1} P_{\mu\nu} P^{\rho\lambda} A_{\rho\lambda}.
\end{align}
Matching with our convention in \cite{Wu1604}, the shear viscous tensor is defined as $\sigma_{\mu\nu} = P_\mu^\rho P_\nu^\sigma \nabla_{(\rho}u_{\sigma)} - \frac1{d-1} P_{\mu\nu} \nabla_\rho u^\rho$ where $\nabla_{(\rho}u_{\sigma)} = \frac12 (\nabla_{\rho} u_{\sigma} + \nabla_{\sigma} u_\rho)$, and the vorticity is defined $\Omega_{\mu\nu} = P_\mu^\rho P_\nu^\lambda \nabla_{[\rho}u_{\lambda]}$ with $\nabla_{[\rho}u_{\lambda]} = \frac12 (\nabla_\rho u_\lambda - \nabla_\lambda u_\rho)$.

(\ref{eq: relativistic fluid 2nd general stress tensor}) has 2 first order transport coefficients (the shear and bulk viscosities $\eta$ and $\zeta$) and 15 second order transport coefficients, of which 9 are coefficients of viscous tensors ($\eta,\eta\tau_\pi,\eta \tau_\pi^*,\kappa,\kappa^*,\lambda_{1,2,3,4}$) and 8 are coefficients of viscous scalars ($\zeta,\zeta\tau_\Pi$, $\xi_{1,2,3,4,5,6}$). A conformal relativistic fluid has only 6 transport coefficients within second order: $\eta,\eta\tau_\pi,\kappa,\lambda_{1,2,3}$. It was shown in \cite{Wu1604} that a relativistic fluid living in curved background can have non-zero $\lambda_4,\xi_{4,5,6}$ and $\kappa,\kappa^*$. The other 9 second order transport coefficients \textit{i.e.}, $\eta\tau_\pi,\eta \tau_\pi^*,\lambda_{1,2,3},\zeta\tau_\Pi,\xi_{1,2,3}$, appear in flat spacetime. The original fluid/gravity correspondence \cite{Bhattacharyya0712} can extract only these 9 transport coefficients.

The study of transport properties within the holographic framework begins with the research on 4d conformal relativistic fluid dual to the AdS$_5$ black hole that comes from trivial dimensional reduction on a near-extremal black D3-brane. The real-time prescription of AdS/CFT correspondence was proposed in \cite{Son0205}, with which Policastro et al. were able to get the shear viscosity \cite{Policastro0205} and the sound wave dispersion relation \cite{Policastro0210} for the 4d conformal fluid. The second order transport coefficients were derived several years later in \cite{Baier0712,Bhattacharyya0712,Kapusta0806,Barnes1004,Arnold1105}. Then the 1st and 2nd order transport coefficients of conformal relativistic fluid in other dimensions \cite{Herzog0210,Natsuume0712,Natsuume0801,VanRaamsdonk0802,Haack0806,Bhattacharyya0809}, and the coupling constant correction to the transport coefficients caused by higher order terms in bulk action \cite{Buchel0406264,Benincasa0510,Buchel0806,Buchel0808NPB,Buchel0808PLB,Brigante0712,Brigante0802,Shaverin1211,Grozdanov1412,
Shaverin1509,Grozdanov2015,Grozdanov1611} were also studied.

The works on nonconformal relativistic hydrodynamics can be divided into two categories, according to the gravity background of the models. The first class is using scalar field deformed AdS$_5$ black holes or the Chamblin-Reall background \cite{Gubser0804,Gubser0806,Li1411,Finazzo1412,Kleinert1610,Bigazzi1006} and their results are usually numerical \cite{Gubser0804,Gubser0806,Li1411,Finazzo1412,Kleinert1610} or approximate \cite{Bigazzi1006}. Thereinto, \cite{Gubser0804,Gubser0806,Li1411,Finazzo1412,Kleinert1610} use the AdS/CFT calculation to study the scalar deformed AdS$_{5}$ black hole and these studies are all numerical, while \cite{Bigazzi1006} adopts the method proposed in \cite{Kanitscheider0901} that the transport coefficients of nonconformal fluid can be deduced from a higher dimensional conformal one and use the result from \cite{Bhattacharyya0809} together with the relations among the 2nd order coefficients from \cite{Romatschke0902} to derive the 2nd order transport coefficients for a 5d Chamblin-Reall background to first order of $\delta=1-3c_s^2$ corrections. The exact results are given in \cite{Kleinert1610}.

The second class of works uses instead brane or supergravity backgrounds from string/M-theory and finds analytic results. Because such backgrounds are 10- or 11-dimensional, they need to be reduced to lower-dimensional manifold in order to use either AdS/CFT calculation or fluid/gravity correspondence to calculate transport coefficients. References of this kind are \cite{Buchel0406200,Benincasa0507,Buchel0812,Buchel0509,Buchel0708,Parnachev0506,Benincasa0605,Mas0703,Natsuume0712,
Natsuume0807,Springer0810,Springer0902,David0901,Kanitscheider0901,Wu1508,Wu1604}, in which \cite{Buchel0406200,Benincasa0507,Buchel0812,Buchel0509,Natsuume0807} are about the relativistic fluid that dual to supergravity backgrounds. And \cite{Parnachev0506,Benincasa0605,Mas0703,Natsuume0712,Natsuume0807,Springer0810,Springer0902,David0901,Kanitscheider0901,
Wu1508,Wu1604} are about NS5-brane \cite{Parnachev0506}, Dp-brane \cite{Mas0703,Natsuume0712,Natsuume0807,Springer0810,Springer0902,David0901,Kanitscheider0901} and compactified D4-brane \cite{Benincasa0605,Wu1508,Wu1604}. To be more specific, \cite{Parnachev0506,Benincasa0605,Mas0703,Springer0810,David0901,Kanitscheider0901,Wu1508} study the first order transport coefficients while \cite{Natsuume0712,Natsuume0807,Springer0902,Wu1604} investigate the second order ones, in which \cite{Natsuume0712,Natsuume0807,Springer0902} calculate the shear and bulk relaxation time $\tau_\pi$ and $\tau_\Pi$ for Dp-brane and \cite{Wu1604} derives all the 7 dynamical second order transport coefficients\footnote{The classification of the second order transport coefficients into dynamical and thermodynamical was proposed in \cite{Moore1210}, according to whether a viscous term contains either $\sigma_{\mu\nu}$ or $\nabla\cdot u$ which can cause entropy increasing.} for compactified D4-brane.

Of the studies reviewed above, only \cite{Finazzo1412,Kleinert1610,Bigazzi1006,Natsuume0712,Natsuume0807,Springer0902,Wu1604}
are about the 2nd order transport coefficients of nonconformal relativistic fluids. Our knowledge about this subject is still very limited, especially for Dp-brane backgrounds. In this paper, we would like to derive all the 7 dynamical second order transport coefficients for Dp-brane with $p=1,2,\cdots,6$, dual to relativistic fluids of dimension $d=p+1=2,3,\cdots,7$, via fluid/gravity correspondence.

It is said that $\lambda_3$ and $\xi_3$ can be derived from fluid/gravity duality \cite{Wu1604}. Since these two coefficients relate with vortical mode of fluid flow, they will only appear in backgrounds which are not isotropic in all spatial directions. This condition can be satisfied in backgrounds with angular momentum \cite{Bhattacharyya0806} or vector charge \cite{Erdmenger0809}, for the vector perturbations from metric tensor (backgrounds with angular momentum) are involved with those from vector current (backgrounds with vector charge). So $\lambda_3$ and $\xi_3$ can not be extracted from branes with $\mathrm{SO}(1,p)$ symmetry in their world-volume directions ($p$ is the spatial dimension of the brane). Thus both of them are 0 for Dp-brane.

The last part of the Introduction is an overview of the paper.

In \cref{sec: dim reduction from 10 to p+2}, we show the procedure of reducing the 10d near-extremal Dp-brane background to a $(p+2)$-dimensional one.
Then we will solve the first order perturbations in \cref{sec: 1st order perturbation}, we find that the tensor perturbations of D5- and D6-brane are non-normalizable in \cref{subsec: D5D6 1st order} thus don't have dual fluid.
So we only need to solve the Dp-brane with $1\leq p \leq 4$ of which the first order perturbations are done in \cref{subsec: Dp 1st order}. Then we calculate the first order stress-energy tensor in \cref{sec: 1st order stress tensor} and find the results agree with previous studies.

The second order perturbations are solved in \cref{sec: 2nd order perturbation}. First we derive the second order expanded metric, the constraint equations of the 2nd order viscous terms and the Navier-Stokes equation in \cref{subsec: 2nd order perturbation general discussion}. Then we redo the calculation for the D3-brane with our convention in \cref{subsec: D3 2nd order perturbations}. The second order perturbations for D4-, D1- and D2-brane are solved in \cref{subsec: solve D4-brane 2nd order perturbation}, \ref{subsec: D1 2nd order} and \ref{subsec: D2 2nd order}, respectively.

\cref{sec: 2nd order stress-energy tensor} will give the final result of the second order stress-energy tensor for Dp-brane. In
\cref{subsec: Dp 2nd order stress-energy tensor} we formulate the final results of the stress-energy tensor and the 2nd order transport coefficients for D4- \cref{eq: D4 2nd order stress-energy tensor}, D1- \cref{eq: D1 2nd order stress-energy tensor} and D2-brane \cref{eq: D2 2nd order stress-energy tensor} both in explicit and unified form \cref{eq: Dp 2nd order stress-energy tensor (general)}. What's more, the gauge theory language of all the transport coefficients are also given in \cref{subsec: Dp 2nd order stress-energy tensor}.
In \cref{subsec: relations of 2nd order coefficients} we talk about the relations satisfied by the 2nd order transport coefficients. The most important observation is that the Romatschke and Kleinert-Probst relations need to be generalized when the fluid is not 4-dimensional. The generalized form can be seen in \cref{eq: relation between xi_1 and lambda_1 (generalized)}, \cref{eq: Romatschke relation curved (generalized)} and \cref{eq: Kleinert-Probst relation (generalized)}.
We give a test on Kanitscheider-Skenderis proposal in \cref{subsec: test of KS proposal} and calculate the dispersion relations in \cref{subsec: dispersion relation}.

\cref{sec: summary} is a summary and some interesting problems for future work will also be mentioned. \cref{app: reduction of bulk metric} is about the dimensional reduction of bulk metric.

\section{Dimensional reduction from 10 to $p+2$ dimensions}\label{sec: dim reduction from 10 to p+2}

This section will deal with the reduction from 10d supergravity of Dp-brane to a $(p+2)$-dimensional effective bulk theory. The 10d action of Dp-brane in Einstein frame is
\begin{align}\label{eq: 10d full action}
  S =&\; \frac1{2\kappa_{10}^2} \int d^{10}x \sqrt{-G} \left[ \mathcal R - \frac12 (\nabla_{\hat M}\phi)^2 - \frac{g_s^2}{2(8-p)!} e^{\frac{p-3}2\phi} F_{8-p}^2 \right] \cr
  & - \frac{1}{\kappa_{10}^2} \int d^9x \sqrt{-H} \mathcal K + \frac{1}{\kappa_{10}^2} \int d^9x \sqrt{-H} \frac{9-p}{2L_p} e^{\frac{3-p}{4(7-p)}\phi}.
\end{align}
The first, second and third part of the above action are separately the bulk term, Gibbons-Hawking surface term and the counter term. The 10d surface gravity is $2\kappa_{10}^2=(2\pi)^7g_s^2l_s^8$ where $g_s$ and $l_s$ are separately the string coupling and the string length. In the bulk term, $G_{\hat M\hat N}$ is the 10d metric with $\mathcal R$ the corresponding Ricci scalar, $\phi$ is the dilaton with zero VEV and $F_{8-p}$ is the magnetic dual of the Ramond-Ramond field minimally coupled with Dp-brane. The 10d Dp-brane background under the near horizon limit that solves the bulk term of (\ref{eq: 10d full action}) is
\begin{align}
  & ds^2 = \left( \frac r{L_p} \right)^\frac{(7-p)^2}8 \Big( -f(r)dt^2 + (dx^i)^2 \Big) + \left( \frac{L_p}r \right)^\frac{(p+1)(7-p)}8 \left( \frac{dr^2}{f(r)} + r^2d\Omega_{8-p}^2 \right), \label{eq: 10d Dp metric} \\
  & e^\phi = \left( \frac r{L_p} \right)^\frac{(p-3)(7-p)}4, \label{eq: dilaton profile} \\
  & F_{\theta_1 \cdots \theta_{8-p}} = g_s^{-1} Q_p \sqrt{\gamma_{8-p}}, \label{eq: RR field}
\end{align}
where $f(r) = 1 - \frac{r_H^{7-p}}{r^{7-p}}$. The parameter $Q_p$ in (\ref{eq: RR field}) is defined as $Q_p = (7-p) L_p^{7-p}$ with
\begin{align}
  L_p^{7-p} = \frac{(2\pi l_s)^{7-p} g_s N}{(7-p) \Omega_{8-p}}, \label{eq: definition of L_p}
\end{align}
where $\Omega_{8-p} = 2\pi^{\frac{9-p}{2}} / \Gamma\left( \frac{9-p}{2} \right)$ is the volume of the $(8-p)$-dimensional unit sphere, and $N$ is the number of the D-branes.

The boundary of (\ref{eq: 10d Dp metric}) is at some constant and large value of $r$ with $H_{\hat M\hat N} = G_{\hat M\hat N} - \mathbf n_{\hat M} \mathbf n_{\hat N}$ the induced metric on it. $\mathbf n_{\hat M}$ is the unit norm in 10d defined as $\mathbf n_{\hat M} = \nabla_{\hat M}r / \sqrt{G^{\hat N\hat P} \nabla_{\hat N}r \nabla_{\hat P}r }$. Then the extrinsic curvature in the Gibbons-Hawking term of (\ref{eq: 10d full action}) is $\mathcal K = -H^{\hat M\hat N} \nabla_{\hat M} \mathbf n_{\hat N}$. The counter term for Dp-brane action can refer to \textit{e.g.}, \cite{Mateos0701}, in which $L_p$ appears in the denominator just to balance the dimension.

Now we rewrite (\ref{eq: 10d Dp metric}) as
\begin{align}\label{eq: reducing form of 10d metric}
  ds^2 = \mathrm{e}^{2\alpha_1 A} g_{MN} dx^M dx^N + L_p^2 \mathrm{e}^{2\alpha_2 A} d\Omega_{8-p}^2
\end{align}
with its induced boundary metric as
\begin{align}\label{eq: reducing form of 9d boundary}
  ds^2 = \mathrm{e}^{2\alpha_1 A} h_{MN} dx^M dx^N + L_p^2 \mathrm{e}^{2\alpha_2 A} d\Omega_{8-p}^2
\end{align}
to reduce the 10d theory described by \cref{eq: 10d Dp metric,eq: 10d full action,eq: dilaton profile,eq: RR field} to a $(p+2)$-dimensional one. The scalar function $A$ can be fixed by comparing (\ref{eq: reducing form of 10d metric}) with (\ref{eq: 10d Dp metric}). In order to make the dilaton $\phi$ not coupled with Ricci scalar in the reduced theory, $\alpha_1,\alpha_2$ should be set as $\alpha_1 = -\frac{8-p}{p},\alpha_2 = 1$. $g_{MN}$ and $h_{MN}$ are the $(p+2)$-dimensional bulk and induced boundary metric, respectively. The coordinate system is $x^{\hat M} = \{x^M,\theta^a\} = \{x^\mu,r,\theta^a\}$ with $x^\mu$, $r$ and $\theta^a$ independent of each other, thus we have $\nabla_{\theta^a} r = 0$. The 10d unit norm reduces to
\begin{align}
  \mathbf n_{\hat M} = {\nabla_{\hat M}r \over \sqrt{ G^{\hat N\hat P} \nabla_{\hat N}r \nabla_{\hat P}r }} = {\nabla_{M}r \over \sqrt{ G^{NP} \nabla_{ N}r \nabla_{P}r }} =  \mathrm{e}^{- \frac{8-p}{p} A} n_M,
\end{align}
where $n_M = \frac{\nabla_M r}{\sqrt{g^{NP} \nabla_N r \nabla_P r}}$ is the unit norm in $p+2$ dimensions. Then one also has
\begin{align}
  \mathcal K = - H^{\hat M\hat N} \nabla_{\hat M} \mathbf n_{\hat N} = \mathrm{e}^{\frac{2(8-p)}{p} A} h^{MN} \nabla_M \left( \mathrm{e}^{- \frac{8-p}{p} A} n_N \right) = \mathrm{e}^{\frac{(8-p)}{p} A} K,
\end{align}
where $K$ is the reduced extrinsic curvature defined as $K = -h^{MN} \nabla_M n_N$. So the full action for $(p+2)$-dimensional reduced bulk theory is
\begin{align}\label{eq: p+2 dim reduced full action}
  S &= \frac{1}{2\kappa_{p+2}^2} \int d ^{p+2}x \sqrt{-g} \left[ R - \frac12 (\partial\phi)^2 - \frac{8(8-p)}{p} (\partial A)^2 + V(\phi,A) \right]  \cr
  & - \frac{1}{\kappa_{p+2}^2} \int d^{p+1}x \sqrt{-h} K + \frac{1}{\kappa_{p+2}^2} \int d^{p+1}x \sqrt{-h} \frac{9-p}{2L_p} e^{- \frac{8-p}{p} A + \frac{3-p}{4(7-p)} \phi}, \cr
  V(\phi,A) &= \frac{(7-p)(8-p)}{L_p^2} \mathrm{e}^{-\frac{16}{p}A} - \frac{Q_p^2}{2L_p^{2(8-p)}} \mathrm{e}^{\frac{p-3}2\phi - \frac{2(p+1)(8-p)}p A},
\end{align}
where $\frac1{2\kappa_{p+2}^2} = \frac{L_p^{8-p} \Omega_{8-p}}{2\kappa_{10}^2}$ is the surface gravity constant in $(p+2)$-dimensional reduced bulk theory.

The reduced background which solves (\ref{eq: p+2 dim reduced full action}) is
\begin{align}\label{eq: p+2 dim reduced background}
  ds^2 &= \left( \frac r{L_p} \right)^\frac{9-p}p (- f(r) dt^2 + (dx^i)^2) + \left( \frac r{L_p} \right)^\frac{p^2-8p+9}p \frac{dr^2}{f(r)}, \cr
  \mathrm{e}^\phi &= \left( \frac r{L_p} \right)^\frac{(p-3)(7-p)}4,~~~~\mathrm{e}^A = \left( \frac r{L_p} \right)^\frac{(p-3)^2}{16}.
\end{align}
Please note that both of the two scalar fields vanish when $p=3$, which is the conformal case that has been solved in \cite{Bhattacharyya0712,Baier0712,Arnold1105,Barnes1004}. The equations of motion (EOM) which can be derived from (\ref{eq: p+2 dim reduced full action}) are the Einstein equation and the EOM of scalar fields $\phi$ and $A$:
\begin{align}
  & E_{MN} - T_{MN} = 0, \label{eq: EOM Einstein eq} \\
  & \nabla^2\phi - \frac{(p-3)(7-p)^2}{4L_p^2} \mathrm{e}^{\frac{p-3}2\phi - \frac{2(p+1)(8-p)}p A} = 0, \label{eq: EOM dilaton} \\
  & \nabla^2 A - \frac{(7-p)}{L_p^2} \mathrm{e}^{-\frac{16}{p}A} + \frac{(p+1)(7-p)^2}{16L_p^2} \mathrm{e}^{\frac{p-3}2\phi - \frac{2(p+1)(8-p)}p A} = 0. \label{eq: EOM A}
\end{align}
where $E_{MN} = R_{MN} - \frac12 g_{MN} R$ is the Einstein tensor and
\begin{align}
  T_{MN} =&\; \frac12 \left( \partial_M \phi \partial_N \phi - \frac12 g_{MN} (\partial \phi)^2 \right) + \frac{8(8-p)}p \left(\partial_M A \partial_N A - \frac12 g_{MN} (\partial A)^2 \right) \cr
  & + \frac12 g_{MN} V
\end{align}
is the energy momentum tensor in $(p+2)$-dimensional spacetime. From the expressions of $\phi$ and $A$ in (\ref{eq: p+2 dim reduced background}) we have $A=\frac{p-3}{4(7-p)} \phi$, which shows that these two scalar fields are not independent thus we only need to solve one of (\ref{eq: EOM dilaton}) and (\ref{eq: EOM A}) to get the scalar perturbations.

\section{The first order perturbations}\label{sec: 1st order perturbation}

We will set $L_p=1$ from now on and restore it when we present our results. Now we switch to the Edington-Finkelstein coordinate by $dt = dv - dr / (r^\frac{7-p}{2} f)$, then the metric in (\ref{eq: p+2 dim reduced background}) after coordinate boost becomes
\begin{align}
  ds^2 = - r^\frac{9-p}p f(r) u_\mu u_\nu dx^\mu dx^\nu + r^\frac{9-p}p P_{\mu\nu} dx^\mu dx^\nu - 2 r^\frac{(p-3)(p-6)}{2p} u_\mu dx^\mu dr.
\end{align}
Following the standard prescription of fluid/gravity correspondence \cite{Bhattacharyya0712}, we promote the boost parameters $u^\mu$ and $r_H$ to be $x^\mu$ dependent and expand them to the first order in $\partial_\mu$ like in \cite{Wu1508}, then we get
\begin{align}\label{eq: 1st order expanded metric}
  ds^2 &= - r^\frac{9-p}p \left( f - \frac{(7-p)r_H^{6-p}}{r^{7-p}}\delta r_H \right) dv^2 + 2 r^\frac{9-p}p (f-1) \delta\beta_i dx^idv \cr
          & + 2 r^\frac{(p-3)(p-6)}{2p} dvdr + r^\frac{9-p}p (dx^i)^2 - 2r^\frac{(p-3)(p-6)}{2p} \delta\beta_i dx^idr.
\end{align}
The general form of perturbations are set to be
\begin{align}\label{eq: general form perturbation ansatz}
  ds^2_{gen.~pert.} &= r^\frac{9-p}p k(x,r) u_\mu(x) u_\nu(x) dx^\mu dx^\nu + 2 r^\frac{9-p}p P_\mu^\rho w_\rho(x,r) u_\nu dx^\mu dx^\nu \cr
  & + r^\frac{9-p}p (\alpha_{\mu\nu}(x,r) + h(x,r) P_{\mu\nu}(x)) dx^\mu dx^\nu - 2 r^\frac{(p-3)(p-6)}{2p} j(x,r) u_\mu dx^\mu dr. \quad
\end{align}
Here we use different conventions for setting the perturbations like $k$ and $w_\mu$ from that of \cite{Wu1508,Wu1604}. After expanded to first order, (\ref{eq: general form perturbation ansatz}) becomes
\begin{align}\label{eq: 1st order perturbation ansatz}
  ds^2_{1st~pert.} &= r^\frac{9-p}p k^{(1)} dv^2 - 2r^\frac{9-p}p w_i^{(1)} dx^idv + r^\frac{9-p}p (\alpha_{ij}^{(1)} + h^{(1)}\delta_{ij})dx^idx^j \cr
 & + 2r^\frac{(p-3)(p-6)}{2p} j^{(1)} dvdr,
\end{align}
in which all the first order perturbations only depend on $r$. This is because the expansion actually happens at a specific point on the boundary (which we set by default the original point). It will apply to all $x^\mu$ after we write the solved metric back into covariant form, according to \cite{Bhattacharyya0712}. (\ref{eq: 1st order perturbation ansatz}) will be solved in this section.

The perturbations in (\ref{eq: 1st order perturbation ansatz}) can be grouped into three independent parts according to the transformation rules under  $\mathrm{SO}(p)$. The EOM can be divided into the dynamical and constraint equations by the original terminology in \cite{Bhattacharyya0712}. The dynamical equations for Dp-brane are
\begin{align}
   E_{ij} - \frac1p \delta_{ij} \delta^{kl} E_{kl} - \left( T_{ij} - \frac1p \delta_{ij} \delta^{kl} T_{kl} \right) &= 0, \label{eq: EOM(ij)} \\
   E_{ri}-T_{ri} &= 0, \label{eq: EOM(ri)} \\
   E_{rr}-T_{rr} &= 0, \label{eq: EOM(rr)} \\
   \nabla^2 \phi - \frac{(p-3)(7-p)^2}{4} e^{\frac{p-3}2 \phi - \frac{2(p+1)(8-p)}p A} &= 0, \label{eq: EOM(phi)}
\end{align}
while the constraint equations are from the vector and scalar sectors:
\begin{align}
  & g^{r0} (E_{0i}-T_{0i}) + g^{rr} (E_{ri}-T_{ri}) = 0, \label{eq: vector constraint} \\
  & g^{rr} (E_{r0} - T_{r0}) + g^{r0} (E_{00} - T_{00}) = 0, \label{eq: scalar constraint 1} \\
  & g^{rr} (E_{rr} - T_{rr}) + g^{r0} (E_{r0} - T_{r0}) = 0, \label{eq: scalar constraint 2}
\end{align}
of which the nature can be specified from their indices. Whether an equation is dynamical or constrained bases on whether it has second derivative order (with respect to $r$) terms of the perturbations: dynamical equations have while the constraint equations do not. Only the EOM of $\phi$ (\ref{eq: EOM(phi)}) is an exception because we do not consider perturbation of $\phi$ in this paper. So (\ref{eq: EOM(phi)}) only contains the first derivative terms of scalar perturbations $h,~j$ and $k$.

Now we put (\ref{eq: 1st order expanded metric}) plus (\ref{eq: 1st order perturbation ansatz}) into the EOM from (\ref{eq: EOM(ij)}) to (\ref{eq: scalar constraint 2}) and get
\begin{align}
  \partial_r (r^{8-p} f(r) \partial_r \alpha_{ij}(r)) + (9-p) r^\frac{7-p}2 \sigma_{ij} &= 0, \label{eq: 1st order EOM(ij) diff} \\
  \partial _r \left( r^{8-p} \partial _r w_i^{(1)} \right) - \frac{9-p}{2} r^\frac{7-p}{2} \partial _0\beta_i &= 0, \label{eq: 1st order EOM(ri) diff} \\
  2p r h'' + p(7-p) h' - 2(9-p) j' &= 0, \label{eq: 1st order EOM(rr) diff} \\
  2(r^{7-p}k)' + 2r^{7-p} f j' + 4(7-p) r^{6-p} j - p r^{7-p} f h' - 2r^\frac{7-p}2 \partial \beta &= 0, \label{eq: 1st order EOM(phi) diff}
\end{align}
which are the dynamical differential equations and
\begin{align}
  & \frac{1}{r_H} \partial _ir_H = - \frac{2}{5-p} \partial _0 \beta_i, \label{eq: 1st order vector constraint diff} \\
  & \frac1{r_H} \partial_0 r_H = - \frac2{9-p} \partial \beta, \label{eq: 1st order scalar constraint 1 diff} \\
  & (r^{7-p} k)' + 2 (7-p) r^{6-p} j - \left( p r^{7-p} - \frac{2p}{9-p} r_H^{7-p} \right) h' - 2 r^\frac{7-p}{2} \partial \beta = 0,
  \label{eq: 1st order scalar constraint 2 diff}
\end{align}
which are the constraint equations. The differential equation from the EOM of $\phi$ (\ref{eq: 1st order EOM(phi) diff}) does not exist for D3-brane. The vector constraint (\ref{eq: 1st order vector constraint diff}) and the first scalar constraint (\ref{eq: 1st order scalar constraint 1 diff}) are only algebraic equations while the second scalar constraint (\ref{eq: 1st order scalar constraint 2 diff}) is a differential one which will be used to solve the scalar perturbations. The constraint equations of the algebraic type exist because the viscous terms are not independent of each other. In first order, one has the following viscous terms:
\begin{align}
  \sigma_{ij} = \partial_{(i}\beta_{j)} - \frac1p \delta_{ij} \partial \beta,~~~~\partial_0 \beta_i,~~~~\partial_i r_H,~~~~\partial \beta,~~~~\partial_0 r_H.
\end{align}
In the above, (\ref{eq: 1st order vector constraint diff}) and (\ref{eq: 1st order scalar constraint 1 diff}) tell us that in vector and scalar part of the viscous tensors, $\partial_i r_H$ and $\partial_0 r_H$ are not independent of $\partial_0 \beta_i$ and $\partial \beta$, respectively. We will only use the latter two in this paper. Since the perturbations are all $\mathrm{SO}(p)$ invariant, one can set $\alpha_{ij} = F(r) \sigma_{ij}$, $w_i = a(r) \partial_0 \beta_i$, $h = F_h(r) \partial \beta,~j = F_j(r) \partial \beta$ and $k = F_k(r) \partial \beta$. Using (\ref{eq: 1st order EOM(rr) diff}), (\ref{eq: 1st order EOM(phi) diff}) and (\ref{eq: 1st order scalar constraint 2 diff}), one can eliminate $j,k$ and get the differential equation for $h$
\begin{align} \label{eq: EOM of h diff}
  \partial_r (r^{8-p} f \partial_r h) + \frac{9-p}p r^\frac{7-p}2 \partial \beta = 0.
\end{align}
Comparing (\ref{eq: EOM of h diff}) with (\ref{eq: 1st order EOM(ij) diff}), one gets
\begin{align}
F_h = \frac1p F, \label{eq: Fh and F}
\end{align}
which reflects the fact that the scalar perturbation $h$ is the trace part of the perturbation tensor at the first order.

Next we are going to solve the EOM for all the Dp-brane in the first order. We will divide our discussion into the unphysical D5 and D6-brane case and the Dp-brane cases with $1\leq p \leq4$. The D3-brane case is actually dual to the renowned $\mathcal N=4$ super Yang-Mills plasma, of which the results are already known in \cite{Bhattacharyya0712,Bhattacharyya0809,Baier0712,Arnold1105}. The reasons to repeat the calculation for D3-brane are: 1). We will use a different way to solve the perturbations at the first order; 2). At the second order, we will use different conventions to define the viscous terms in order to make them applicable for various dimensions and more convenient to transfer to covariant form. As a result, some of the intermediate results will be different from \cite{Bhattacharyya0712}.

\subsection{The D5 and D6-brane case}\label{subsec: D5D6 1st order}

Solving the tensor perturbation equation (\ref{eq: 1st order EOM(ij) diff}) for $p$ = 5 and 6, we get
\begin{align}
  F(r) &= C_2 - \ln(r^2 - r_H^2) + \frac{C_1}{2r_H^2} \ln\left( 1 - \frac{r_H^2}{r^2} \right), \quad \text{(D5-brane)} \label{eq: D5 F(r)} \\
  F(r) &= C_2 - 4r^{\frac12} + 4r_H^{\frac12} \,{\rm artanh}\, \sqrt{\frac{r}{r_H}} + \frac{C_1}{r_H} \ln\left( 1 - \frac{r_H}{r} \right), \quad \text{(D6-brane)} \label{eq: D6 F(r)}
\end{align}
where $C_{1,2}$ are the integration constants. The function ``$\,{\rm artanh}\,$" is sometimes misnamed as ``$\,{\rm arctanh}\,$", the ``ar" in artanh means area not arc.

Problems arise in \cref{eq: D5 F(r),eq: D6 F(r)}. One can see that both of these two solutions diverge at $r \to \infty$ thus do not correspond to physical mode. From $F_h = \frac1p F$ we can conclude that we will not have a physical solution for $h$, either. This means D5 and D6-brane are not dual to any physical relativistic fluid. Similar conclusions have also been made in \cite{Natsuume0807}. The author finds the formula for deriving the shear relaxation time (originally proposed in \cite{Kapusta0806}) vanishes for D5-brane and diverges for D6-brane. The physical reason behind this is also pointed out in \cite{Natsuume0807} that Dp-branes for $p\geq 5$ have instability: the sound speed vanishes for D5-brane and become negative for D6-brane. We will see this instability by calculating the sound speed in \cref{sec: 1st order stress tensor}.

The fact that D5 and D6-brane are not dual to any physical relativistic fluid will save us plenty of work, we only need to consider the Dp-brane with ${1\leq p \leq 4}$ from now on.

\subsection{The Dp-brane case with $1\leq p \leq 4$}\label{subsec: Dp 1st order}

We can solve the tensor perturbations for Dp-brane with $2\leq p\leq 4$ from (\ref{eq: 1st order EOM(ij) diff}) as
\begin{align}
  F(r) =&\; \frac2{5r_H^{3/2}} \Bigg[ 2\sin\frac{2\pi}5 \arctan{ 2\sin\frac\pi5 \sqrt{rr_H}\over r - r_H} - 2\sin\frac\pi5 \arctan{ 2\sin\frac{2\pi}5 \sqrt{rr_H} \over r-r_H} \cr
  & - 2\cos\frac{2\pi}5 \,{\rm artanh}\, \frac{2\cos\frac{\pi}5 \sqrt{rr_H}}{r+r_H} - 2\cos\frac\pi5 \,{\rm artanh}\,\frac{2\cos\frac{2\pi}5 \sqrt{rr_H}}{r+r_H}  \cr
  &+ \ln\frac{(\sqrt r + \sqrt{r_H})^2 (r^4 + r^3r_H + r^2r_H^2 + rr_H^3 + r_H^4)}{r^5} \Bigg], \quad \text{(D2-brane)} \label{eq: D2 F(r)} \\
  F(r) =&\; \frac1{2r_H} \left[ 2\arctan\frac{r_H}{r} + \ln\frac{(r + r_H)^2 (r^2 + r_H^2)}{r^4} \right], \quad \text{(D3-brane)} \label{eq: D3 F(r)} \\
  F(r) =&\; \frac1{3r_H^{1/2}} \left[ 2\sqrt3 \arctan\frac{ \sqrt{3r r_H} }{r - r_H} \right. \cr
  & \left. + \ln{ (\sqrt r + \sqrt r_H)^4 (r + \sqrt{rr_H} + r_H)^2 (r^2 + r r_H + r_H^2) \over r^6 } \right], \quad \text{(D4-brane)} \label{eq: D4 F(r)}
\end{align}
Note that D1-brane only has 1 spatial dimension thus can not support tensor perturbation. $F(r)$ for D2-brane is more complicated than the other two cases, yet it can still be cast into a neat form as (\ref{eq: D2 F(r)}). We write the first order tensor perturbation of D3-brane \textit{i.e.}, (\ref{eq: D3 F(r)}) in a different form from that in \cite{Bhattacharyya0712} by using the identity
\begin{align}\label{eq: arctan identity}
\frac\pi2 - \arctan x = \arctan \frac1x,
\end{align}
which is easy to verify.

The D4-brane tensor perturbation (\ref{eq: D4 F(r)}) is the same as in the compactified case \cite{Wu1508,Wu1604} which can be shown by first using (\ref{eq: arctan identity}) and then the arctangent addition formula. So one can draw the conclusion that compactifying one direction of D4-brane will not effect the coefficients of viscous tensors, \textit{i.e.}, $\eta,\tau_\pi$ and $\lambda_{1,2}$ will not change compared with the compactified D4-brane case. This can also be seen from the differential equation of second order tensor perturbation for D4-brane later.

The vector perturbations for Dp-brane with $1\leq p \leq 4$ can be solved from (\ref{eq: 1st order EOM(ri) diff}) as $w_i^{(1)}(r) = a(r) \partial _0\beta_i$ where we define
\begin{align}
  a(r) = -\frac{2}{(5-p) r^\frac{5-p}2}.
\end{align}
From the above we can see that the vector perturbation diverges for $p=5$ and can not be normalized for $p=6$. Since we use different convention for setting the vector perturbation in (\ref{eq: general form perturbation ansatz}), that's why the above for $p=4$ is not the same as in \cite{Wu1508}. The advantage of this convention is that it can circumvent the appearance of divergent part at the second order, compared with \cite{Bhattacharyya0712,Wu1508,Wu1604}. The vector viscous term $\partial _0\beta_i$ for D1-brane will reduce to a scalar $\partial _0\beta_1$ but we will still call it a vector viscous term throughout the paper.

In solving the first order scalar perturbations of Dp-brane, we have some problems on D3-brane case. As it has been pointed out that the EOM of dilaton (\ref{eq: EOM(phi)}) does not exist in D3-brane case and the scalar perturbations need to be solved under the ``background field" gauge \cite{Bhattacharyya0712}. Actually, choosing a gauge condition is just adding one more equation factitiously. We also know that the scalar perturbations of D3-brane are trivial, which indicates that they will not be written in a compact form together with the other Dp-branes. In order to express the scalar perturbation solutions of Dp-brane in a unified form, we will use the condition (\ref{eq: Fh and F}) for D3-brane. Thus the 3 scalar perturbations of D3-brane can be solved from (\ref{eq: 1st order EOM(rr) diff}), (\ref{eq: 1st order scalar constraint 2 diff}) together with (\ref{eq: Fh and F}).

The other cases can be solved from (\ref{eq: 1st order EOM(rr) diff}), (\ref{eq: 1st order EOM(phi) diff}) and (\ref{eq: 1st order scalar constraint 2 diff}). Then the 1st order scalar perturbations of Dp-brane with $1\leq p\leq 4$ can be nicely formulated as
\begin{align}
  F_h &= \frac1p F, \quad F_j = - \frac{2}{9-p} \frac{ r^\frac{9-p}{2} - r_H^\frac{9-p}{2} }{ r^{7-p} - r_H^{7-p} } + \frac{5-p}{2(9-p)} F, \cr
  F_k &= \frac4{ (9-p) r^\frac{5-p}{2} } - \frac{1}{9-p} \left( 5-p + \frac{2r_H^{7-p}}{r^{7-p}} \right) F.
  \label{eq: Dp 1st order hjk solutions}
\end{align}
In D1-brane case $F$ is just $F_h$ and we only need to keep in mind that for $p=1$ $F$ has nothing to do with tensor perturbation. Thus $F_j$ and $F_k$ of D1-brane relate directly to $F_h$ which is
\begin{align}
  F_h(r) = \frac1{6r_H^2} \left[ 2\sqrt3 \arctan\frac{\sqrt3 r_H^2}{2r^2+r_H^2} + 3\ln\frac{r^4 + r^2 r_H^2 + r_H^4}{r^4} \right].
\end{align}
Compared with the solution in compactified D4-brane \cite{Wu1508,Wu1604}, $F_j$ and $F_k$ do not change while $F_h$ does\footnote{$F_k$ here is $\frac{1}{r^3}$ times the $F_k$ in \cite{Wu1508,Wu1604} due to the convention change on the perturbation ansatz (\ref{eq: general form perturbation ansatz}).}. This is because $h$ relates with the spatial trace part of the perturbation ansatz, as can be seen from (\ref{eq: 1st order perturbation ansatz}). And the spatial dimension of compactified D4-brane is 3 while it is 4 for D4-brane case here.

\section{The first order stress-energy tensor}\label{sec: 1st order stress tensor}

The boundary stress-energy tensor is defined as the large $r$ limit of the following Brown-York tensor
\begin{align}\label{eq: Dp-brane Brown-York tensor}
  T_{\mu\nu} = \frac{1}{\kappa_{p+2}^2} \lim_{r\to\infty} \left( \frac r{L_p} \right)^{\frac{(9-p)(p-1)}{2p}} \left( K_{\mu\nu} - h_{\mu\nu}K - \frac{9-p}{2{L_p}} \left( \frac r{L_p} \right)^{-\frac{(p-3)^2}{2p}} h_{\mu\nu} \right),
\end{align}
from which one can calculate the first order stress-energy tensor for Dp-brane:
\begin{align}\label{eq: 1st order stress-energy tensor}
  T_{\mu\nu} = \frac{1}{2 \kappa_{p+2}^2} \left[ {r_H^{7-p} \over L_p^{8-p}} \left( \frac{9-p}{2} u_\mu u_\nu + \frac{5-p}{2} P_{\mu\nu} \right) - \left( \frac{r_H}{L_p} \right) ^\frac{9-p}{2} \left( 2\sigma_{\mu\nu} + \frac{2(p-3)^2}{p(9-p)} P_{\mu\nu} \partial u \right) \right].
\end{align}
One can extract the energy density, pressure and first order transport coefficients from (\ref{eq: 1st order stress-energy tensor}) as:
\begin{align}\label{eq: Dp 1st order transport coefficients}
  \varepsilon &= \frac{1}{2 \kappa_{p+2}^2} \frac{9-p}{2} {r_H^{7-p} \over L_p^{8-p}}, \qquad \mathfrak p = \frac{1}{2 \kappa_{p+2}^2} \frac{5-p}{2} {r_H^{7-p} \over L_p^{8-p}}, \cr
   \eta &= \frac{1}{2 \kappa_{p+2}^2} \left( \frac{r_H}{L_p} \right)^\frac{9-p}{2}, \qquad \zeta = \frac{1}{2 \kappa_{p+2}^2} \frac{2(p-3)^2}{p(9-p)} \left( \frac{r_H}{L_p} \right)^\frac{9-p}{2}.
\end{align}
From (\ref{eq: p+2 dim reduced background}) one gets the Hawking temperature
\begin{align}
  T = {(7-p) r_H^\frac{5-p}{2} \over 4\pi L_p^\frac{7-p}{2}}, \label{eq: Hawking temperature of Dp}
\end{align}
thus the entropy is
\begin{align}
  s = \frac{\varepsilon+\mathfrak p}{T} = \frac{1}{2 \kappa_{p+2}^2} 4\pi \left( \frac{r_H}{L_p} \right)^\frac{9-p}{2}.
\end{align}
Then we have
\begin{align}
  c_s^2 = \left( \frac{\partial \mathfrak p}{\partial \varepsilon} \right)_s = \frac{5-p}{9-p}, \qquad \frac\eta s = \frac{1}{4\pi}, \qquad \frac\zeta\eta = \frac{2(p-3)^2}{p(9-p)}.
\end{align}
The above results agree with that of \cite{Kanitscheider0901,David0901,Mas0703,Springer0810}. One can see from the sound speed that D5 and D6-brane are indeed unphysical since the sound speed separately become zero and negative for $p=5$ and 6.

\section{The second order perturbations}\label{sec: 2nd order perturbation}

\subsection{General discussions}\label{subsec: 2nd order perturbation general discussion}

In this section we are going to solve the second order perturbations for Dp-brane with $1\leq p\leq 4$. The start line is the first order covariant metric valid on the whole boundary:
\begin{align}\label{eq: 1st order full metric}
  ds^2 =& -r^\frac{9-p}{p} [ f(r_H(x),r) - k(r_H(x),r) ] u_\mu(x) u_\nu(x) dx^\mu dx^\nu \cr
  &+ 2 r^\frac{9-p}{p} P_\mu^\rho(x,r) w_\rho(x,r) u_\nu(x) dx^\mu dx^\nu  \cr
  & + r^\frac{9-p}{p} [ P_{\mu\nu}(x,r) + \alpha _{\mu\nu}(r_H(x),r) + h(r_H(x),r) P_{\mu\nu}(x,r) ] dx^\mu dx^\nu \cr
  & - 2 r^\frac{(p-3)(p-6)}{2p} [ 1+j(r_H(x),r) ] u_\mu(x) dx^\mu dr.
\end{align}
Then expand $r_H(x)$ and $\beta_i(x)$ to second derivative order like in \cite{Wu1604}. If we use $\mathcal{F}$ to denote any of $F, F_h, F_j$ and
$F_k$, then $\mathcal{F}$ should be expanded like
\begin{align}
  \mathcal{F}(r_H(x),r) = \mathcal{F}(r) + \delta \mathcal F(r_H(x),r) = \mathcal{F}(r) - \frac{(5-p) \mathcal{F}(r) + 2r \mathcal{F}'(r)}{2r_H} \delta r_H.
\end{align}
Here $\mathcal F(r)$ is $\mathcal{F}(r_H(x),r)$ at the original point $x^\mu =0$ and $\delta r_H = x^\mu \partial_\mu r_H$. $\delta \mathcal F(r_H(x),r)$ will be denoted shortly as $\delta\mathcal F$ in the following. Thus (\ref{eq: 1st order full metric}) should be expanded as
\begin{small}
\begin{align}\label{eq: 2nd order expanded metric Dp}
  ds^2 =& -r^\frac{9-p}{p} \bigg[ f + (f-1)\delta\beta_i\delta\beta_i - \frac{(7-p)r_H^{6-p}}{r^{7-p}}\delta r_H - \frac{(7-p)r_H^{6-p}}{2r^{7-p}} \delta^2r_H - \frac{(7-p)r_H^{6-p}}{r^{7-p}}\delta r_H^{(1)} \cr
  & - \frac{(7-p)(6-p) r_H^{5-p}}{2r^{7-p}}(\delta r_H)^2 - (F_k+\delta F_k)\partial \beta - F_k(\delta\partial \beta + \delta\beta_i \partial_0\beta_i) - 2 a(r)\delta\beta_i \partial_0\beta_i \bigg] dv^2 \cr
  & + 2r^\frac{9-p}{p} \bigg[ (f-1)(\delta\beta_i + \frac12 \delta^2\beta_i) - a(\partial_0\beta_i + \delta\partial_0\beta_i + \delta\beta_j\partial_j\beta_i) - \frac{(7-p)r_H^{6-p}}{r^{7-p}}\delta r_H \delta\beta_i \cr
  & - F_k \partial \beta \delta\beta_i - F \delta\beta_j \partial_{(i}\beta_{j)} \bigg] dvdx^i \cr
  & + 2r^\frac{(p-3)(p-6)}{2p} \bigg[ 1 + (F_j+\delta F_j)\partial \beta + F_j(\delta\partial \beta + \delta\beta_i\partial_0\beta_i) + \frac12 \delta\beta_i\delta\beta_i \bigg] dvdr \cr
  & + r^\frac{9-p}{p} \bigg[ \delta_{ij} + (1-f)\delta\beta_i\delta\beta_j + 2a \delta\beta_{(i}\partial_{|0|}\beta_{j)} + (F + \delta F) \partial_{(i}\beta_{j)} \cr
  & + F \left( \delta\partial_{(i} \beta_{j)} + \delta\beta_{(i} \partial_{|0|}\beta_{j)} \right) \bigg] dx^idx^j - 2 r^\frac{(p-3)(p-6)}{2p} \bigg( \delta\beta_i + \frac12 \delta^2\beta_i + F_j \partial\beta \delta\beta_i \bigg) dx^i dr.
\end{align}
\end{small}
In the above $\delta\beta_i = x^\mu \partial_\mu \beta_i$, similarly as $\delta r_H$. We also define $\delta^2\beta_i = x^\mu x^\nu \partial_\mu \partial_\nu \beta_i$ and $\delta\beta_{(i} \partial_{|0|}\beta_{j)} = \frac12 (\delta\beta_{i} \partial_{0}\beta_{j} + \delta\beta_{j} \partial_{0}\beta_{i})$. Note after the expansion with respect to the boundary derivative terms, the metric becomes a quadratic function in boundary coordinates $x^\mu$ and it is also a complicate function of $r$ via $F$, $F_j$ and $F_k$. We will set $r_H=1$ hereafter and restore it when we give the results of stress-energy tensors.

Since Einstein equation is second order in derivatives, thus one will get differential equations with many second order derivatives of $r_H$ and $\beta_i$ which can be written as the spatial 2nd order viscous terms. We list them in \cref{tab: Dp 2nd order spatial viscous tensors} where $\Omega_{ij} = \partial_{[i} \beta_{j]}$ is the spatial component of $\Omega_{\mu\nu}$. Some of the conventions for the 2nd order spatial viscous terms in this paper are different from \cite{Bhattacharyya0712,Wu1604} because we need to make them appropriate for general value of $p$, not only for the case of $p=3$. Since the viscous terms of D1-brane are very different from the other cases, thus we will list them separately when we talk about D1-brane later.

We would like to stress the changes of \cref{tab: Dp 2nd order spatial viscous tensors} from that in \cite{Bhattacharyya0712,Wu1604}:
First, we rewrite some terms with $l_i = \epsilon_{ijk} \Omega_{ij}$ in terms of $\Omega_{ij}$ since $l_i$ only exists for $p=3$ while $\Omega_{ij}$ exists in all cases of $p\geq 2$, \textit{e.g.} $l_il_i \to \Omega_{ij}^2$ and $l_il_j - \frac1p \delta_{ij} l_k^2 \to \Omega_i^{~k}\Omega_{jk} - \frac1p \delta_{ij} \Omega_{kl}^2$.
Second, we remove $\mathfrak S_2=l_i\partial_0\beta_i$ and $\mathbf{v}_{3i}=\partial_0 l_i$ since they both contain $l_i$ and can not be rewrite as  scalar and vector with the role of $l_i$ replaced by $\Omega_{ij}$. In order to be convenient to compare with previous studies, we will not change the ordinal numbers of the other $\mathrm{SO}(p)$ invariant scalars and vectors.
Third, we redefine $\mathbf{t}_{2}$ and $\mathfrak T_{2,3}$ of $\mathrm{SO}(p)$ tensors in terms of $\Omega_{ij}$ to express the antisymmetric tensor constraint, \textit{i.e.} (\ref{eq: 2nd order constraint antisym tensor}) for general $p$. Thus the viscous tensors in \cref{tab: Dp 2nd order spatial viscous tensors} do not only contain the traceless-symmetric tensors but also the antisymmetric tensors. This is different from that in \cite{Bhattacharyya0712,Wu1604}. We will suppress the vector and tensor indices of the spatial viscous terms from now on.

\begin{table}
\centering
\begin{tabular}{|l|l|l|}
  \hline
  Scalars of $\mathrm{SO}(p)$ & Vectors of $\mathrm{SO}(p)$ &\quad \quad  Tensors of $\mathrm{SO}(p)$ \\ \hline\hline
 $\mathbf{s}_1=\frac1{r_H}\partial_0^2r_H$ & $\mathbf{v}_{1i} = \frac1{r_H} \partial_0\partial_i r_H$ & $\mathbf{t}_{1ij} = \frac1{r_H} \partial_i\partial_j r_H - \frac1p \delta_{ij} \mathbf{s}_3$ \\
  $\mathbf{s}_2 = \partial_0\partial_i\beta_i$ & $\mathbf{v}_{2i} = \partial_0^2\beta_i$ & $\mathbf{t}_{2ij} = \partial_0 \Omega_{ij}$ \\
  $\mathbf{s}_3 = \frac1{r_H}\partial_i^2r_H$ & $\mathbf{v}_{4i}=\partial_j\Omega_{ij}$  & $\mathbf{t}_{3ij} = \partial_0\sigma_{ij}$ \\
  $\mathfrak S_1 = \partial_0\beta_i\partial_0\beta_i$ & $\mathbf{v}_{5i} = \partial_j\sigma_{ij}$  & $\mathfrak T_{1ij} = \partial_0\beta_i\partial_0\beta_j - \frac1p \delta_{ij} \mathfrak S_1$ \\
  $\mathfrak S_3 = (\partial_i\beta_i)^2$ &   $\mathfrak V_{1i} = \partial_0\beta_i\partial\beta$   & $\mathfrak T_{2ij} = \sigma_{[i}^{~~k} \Omega_{j]k}$ \\
  $\mathfrak S_4 = \Omega_{ij} \Omega_{ij}$ & $\mathfrak V_{2i} = \partial_0\beta_j \Omega_{ij}$  & $\mathfrak T_{3ij} = \Omega_{ij} \partial \beta$ \\
  $\mathfrak S_5=\sigma_{ij}\sigma_{ij}$ &  $\mathfrak V_{3i} = \partial_0\beta_j \sigma_{ij}$  & $\mathfrak T_{4ij}=\sigma_{ij}\partial\beta$ \\
                                     &                                                            & $\mathfrak T_{5ij} = \Omega_i^{~k}\Omega_{jk} - \frac1p \delta_{ij} \mathfrak S_4$ \\
                                     &                                                            & $\mathfrak T_{6ij} = \sigma_i^{~k}\sigma_{jk} - \frac1p \delta_{ij} \mathfrak S_5$ \\
                                     &                                                            & $\mathfrak T_{7ij} = \sigma_{(i}^{~~k} \Omega_{j)k}$ \\
\hline
\end{tabular}
\caption{\label{tab: Dp 2nd order spatial viscous tensors} A list of second derivative order $\mathrm{SO}(p)$ invariant viscous terms for $p \geq 2$. $\mathfrak T_{2,5,6}$ are zero when $p=2$.}
\end{table}

The viscous terms listed in \cref{tab: Dp 2nd order spatial viscous tensors} satisfy 6 constraint equations which are the 2nd order counterparts of \cref{eq: 1st order vector constraint diff,eq: 1st order scalar constraint 1 diff}. The usage of them is in deriving the Navier-Stokes equation later. These 6 constraints can be solved from $\partial_\mu \partial^\rho T^{(0)}_{\rho\nu} = 0$ by expanding it to second derivative order, the results can be listed as
\begin{align}
  \frac{9-p}{2} \frac{1}{r_H} \partial _0^2 r_H + \partial_0\partial \beta - \frac{2}{9-p} (\partial \beta)^2 - \frac{4}{5-p} \partial_0\beta_i \partial_0\beta_i = 0, \label{eq: 2nd order constraint (00) raw}  \\
  \frac{5-p}{2} \frac{1}{r_H} \partial_i^2 r_H + \partial_0 \partial \beta - \frac{2}{5-p} \partial_0\beta_i \partial_0\beta_i - \frac{5-p}{9-p} (\partial \beta)^2 + \partial_i \beta_j \partial _j\beta_i = 0, \label{eq: 2nd order constraint (ii) raw} \\
  \frac{5-p}{2} \frac{1}{r_H} \partial _0\partial _i r_H + \partial _0^2\beta_i - \frac{7-p}{9-p} \partial _0\beta_i \partial \beta + \partial_0 \beta_j \partial _j\beta_i = 0, \label{eq: 2nd order constraint (0i) raw} \\
  \frac{9-p}{2} \frac{1}{r_H} \partial _0\partial _i r_H + \partial_i \partial \beta - \frac{2}{5-p} \partial _0\beta_i \partial \beta - \frac{4}{5-p} \partial_0 \beta_j \partial _i\beta_j = 0, \label{eq: 2nd order constraint (i0) raw} \\
  \partial _0 \Omega_{ij} - \frac{5-p}{9-p} \Omega_{ij} \partial \beta - \partial_k\beta_{[i} \partial_{j]}\beta_k = 0, \label{eq: 2nd order constraint antisym tensor raw} \\
  \frac{5-p}{2} \frac{1}{r_H} \partial_i \partial _j r_H + \partial _0 \partial_{(i}\beta_{j)} - \frac{2}{5-p} \partial_0\beta_i \partial_0\beta_j - \frac{5-p}{9-p} \partial_{(i}\beta_{j)} \partial\beta + \partial _k\beta _{(i} \partial _{j)} \beta _k = 0. \label{eq: 2nd order constraint (ij) raw}
\end{align}
Here $T^{(0)}_{\mu\nu}$ is the zeroth derivative order of (\ref{eq: 1st order stress-energy tensor}), \textit{i.e.} the thermodynamic part of the 1st order stress-energy tensor of the relativistic fluid. In terms of the viscous $\mathrm{SO}(p)$ invariant terms, the above can be reformulated as
\begin{align}
  & \mathbf s_1 + \frac2{9-p} \mathbf s_2 - \frac{8}{(9-p)(5-p)} \mathfrak S_1 - \frac{4}{(9-p)^2} \mathfrak S_3 = 0, \label{eq: 2nd order constraint (00)} \\
  & \mathbf s_2 + \frac{5-p}{2} \mathbf s_3 - \frac{2}{5-p} \mathfrak S_1 + \frac{(p-3)^2}{p(9-p)} \mathfrak S_3 - \mathfrak
  S_4 + \mathfrak S_5 = 0, \label{eq: 2nd order constraint (ii)} \\
  & \mathbf v_1 + \frac{2}{5-p} \mathbf v_2 + \frac{2(p^2-8p+9)}{p(9-p)(5-p)} \mathfrak V_1 - \frac{2}{5-p} \mathfrak V_2 + \frac{2}{5-p} \mathfrak V_3 = 0, \label{eq: 2nd order constraint (0i)} \\
  & \mathbf v_1 + \frac{2p (\mathbf v_4 + \mathbf v_5)}{(9-p)(p-1)} - \frac{4(p+2)}{p(9-p)(5-p)} \mathfrak V_1 - \frac{8 (\mathfrak V_2 + \mathfrak V_3)}{(9-p)(5-p)} = 0, \label{eq: 2nd order constraint (i0)} \\
  & \mathbf t_2 - 2 \mathfrak T_2 + \frac{p^2 - 7p + 18}{p(9-p)} \mathfrak T_3 = 0, \label{eq: 2nd order constraint antisym tensor} \\
  & \mathbf t_1 + \frac{2}{5-p} \mathbf t_3 - \frac{4}{(5-p)^2} \mathfrak T_1 + \frac{2(p^2-7p+18)}{p(9-p)(5-p)} \mathfrak T_4 - \frac{2}{5-p} \mathfrak T_5 + \frac{2}{5-p} \mathfrak T_6 = 0. \label{eq: 2nd order constraint (ij)}
\end{align}
The above 6 constraints from (\ref{eq: 2nd order constraint (00)}) to (\ref{eq: 2nd order constraint (ij)}) are completely appropriate for D3 and D4-brane. Since all the components of $\mathfrak T_2$, $\mathfrak T_5$ and $\mathfrak T_6$ are 0 for $p=2$ thus we only need to remove them from (\ref{eq: 2nd order constraint antisym tensor}) and (\ref{eq: 2nd order constraint (ij)}) to get the constraints for D2-brane. The D1-brane needs more changes: First, it does not have tensor part so there should be no tensorial constraints; Next, we need to remove the terms $\mathfrak V_{2,3}$ from the vector constraint and $\mathfrak S_{4,5}$ from scalar constraint because these terms consist of $\sigma_{ij}$ and $\Omega_{ij}$; The last is to use $\mathbf v_4 + \mathbf v_5 = \frac{p-1}{p} \partial_i \partial \beta$ to rewrite (\ref{eq: 2nd order constraint (i0)}) since in this way there is no $p-1$ in the denominator which will blow up as $p=1$. The second order constraints for D1-brane will be given later.

It is interesting to compare the above constraints with that of the compactified D4-brane \cite{Wu1604} and the D3-brane \cite{Bhattacharyya0712}. Since the convention for the $\mathrm{SO}(p)$ invariant viscous terms are different from that of \cite{Bhattacharyya0712}, thus the coefficients of the viscous terms here will also change for D3-brane. Also note that the antisymmetric part of the tensor constraint (\ref{eq: 2nd order constraint antisym tensor}) is  tensorial here but it is of axial vector type in the cases of D3-brane \cite{Bhattacharyya0712} and compactified D4-brane \cite{Wu1604}. Since the spatial dimension for both of these two cases are 3 in which one can define the axial vector $l_i$ out of $\Omega_{ij}$, but this can not be done in other dimensions. Thus we write the antisymmetric tensor constraint in terms of three newly defined antisymmetric viscous tensors\footnote{ The old definitions of $\mathbf t_2$ and $\mathfrak T_{2,3}$ contain $l_i$ which are not used in calculation in \cite{Bhattacharyya0712,Wu1508,Wu1604}. }: $\mathbf t_2 = \partial_0 \Omega_{ij}$, $\mathfrak T_2 = \sigma_{[i}^{~k} \Omega_{j]k}$ and $\mathfrak T_3 = \Omega_{ij} \partial \beta$.

There are also two other important equations which can be derived from the full first order stress-energy tensor (\ref{eq: 1st order stress-energy tensor}), which we denote as $T^{(0+1)}_{\mu\nu}$. They are the Navier-Stokes equations $\partial^\mu T^{(0+1)}_{\mu\nu} = 0$ with the index $\nu$ being set to 0 and $i$:
\begin{align}
  \frac{1}{r_H^{(p-3)/2}} \partial_0 r_H^{(1)} =&\; \frac{4(p-3)^2}{p(9-p)^2(7-p)} \mathfrak S_3 + \frac{4}{(9-p)(7-p)} \mathfrak S_5, \label{eq: Navier-Stokes 0}\\
  \frac{1}{r_H^{(p-3)/2}} \partial_i r_H^{(1)} =&\; \frac{4(p-3)^2\,\mathbf v_4 + 16p \, \mathbf v_5}{(9-p)(7-p)(5-p)(p-1)}
  + \frac{2(p-1)(p^2 - 22p + 77)}{p(9-p)(7-p)(5-p)^2} \mathfrak V_1 \cr
 & - \frac{2(19 - 3p)}{(9-p)(7-p)(5-p)} \mathfrak V_2 - \frac{2(p^2 - 14p + 77)}{(9-p)(7-p)(5-p)^2} \mathfrak V_3. \label{eq: Navier-Stokes i}
\end{align}
The above two equations are suitable for $2\leq p \leq 4$. As what we have specified for the 2nd order constraints of D1-brane case, we need to remove $\mathfrak S_5$ and $\mathfrak V_{2,3}$ first and then recast the terms with $\mathbf v_{4,5}$ into $\partial_1^2 \beta_1$. The explicit form of the Navier-Stokes equations for $p=1$ will be offered later. As we have said that in deriving these two equations, we need to use the constraints from (\ref{eq: 2nd order constraint (00) raw}) to (\ref{eq: 2nd order constraint (ij) raw}). The Navier-Stokes equations will be used in deriving the differential equations for the 2nd order perturbations. They can also be got from the bulk vector and scalar constraint equations at second order, which will be specified later.

Following \cite{Bhattacharyya0712,Wu1604}, the second order perturbations can be solved by integrations as
\begin{align}
  \alpha_{ij}^{(2)}~\text{or}~h^{(2)} &= \int_{r}^{\infty} \frac{-1}{x^{8-p} f}dx \int_{1}^{x} S_{\alpha,h}(y) dy, \label{eq: integrate to get alpha&h} \\
  j^{(2)}~\text{or}~k^{(2)} &= - \int_r^\infty S_{j,k}(x) dx, \label{eq: integrate to get j&k} \\
  w_{i}^{(2)} &= \int_{r}^{\infty} \frac{1}{x^{8-p}} dx \int_{x}^{\infty} S_w(y) dy, \label{eq: integrate to get w}
\end{align}
where $S_{\#}$ with $\# = \{\alpha,w,h,j,k\}$ are the source terms which can be easily understood as the right hand side of the perturbation equations. The above integrations are suitable for the nonconformal branes, \textit{i.e.}, $p=1,2,4$. For $p=3$, $h^{(2)}$ needs to be integrated in the same way as $w_{i}^{(2)}$ but not like $\alpha_{ij}^{(2)}$ since the scalar part of D3-brane is trivial. While the other perturbations of D3-brane are still integrated in the same way as the other Dp-branes.

\subsection{The D3-brane case}\label{subsec: D3 2nd order perturbations}

Although the second order transport coefficients of D3-brane has been studied in \cite{Bhattacharyya0712}, we still would like to redo the calculation here. The reason is that we use different conventions, thus the intermediate steps of the calculation will also change. Further more, the D3-brane calculation can help us to understand the nonconformal cases.

At the second order there is no simple relation between $h$ and $\alpha_{ij}$ like (\ref{eq: Fh and F}). Since in D3-brane case we do not have the dilaton equation and one only has (\ref{eq: EOM(rr)}) and (\ref{eq: scalar constraint 2}) at hand to solve $h,~j$ and $k$, which are not enough. Thus we need a gauge condition to eliminate one of the degrees of freedom. Here we use the ``background field" gauge $g^{(0)MN}g_{MN}^{(n)} = 0$ from \cite{Bhattacharyya0712}, which gives $j^{(n)} = - \frac32 h^{(n)}$, where $(n)$ specifies the order of each perturbation. With this we can solve $h,j,k$ for D3-brane at first order as
\begin{align}
  h ^{(1)} = j^{(1)}=0, \quad k^{(1)} = \frac2{3r} \partial \beta.
\end{align}

Since $F_{h,j,k}$ are trivial for D3-brane under the background field gauge, the second order expanded metric of D3-brane can not be got by simply setting $p=3$ in (\ref{eq: 2nd order expanded metric Dp}), but needs to be derived separately. The result is:
\begin{align}\label{eq: 2nd order expanded metric D3}
  ds^2 =& - r^2 \bigg[ f + (f-1)\delta\beta_i\delta\beta_i - \frac{4r_H^3}{r^4}\delta r_H - \frac{2r_H^3}{r^4}\delta^2 r_H - \frac{4r_H^3}{r^4}\delta r_H^{(1)} - \frac{6r_H^2}{r^4}(\delta r_H)^2 \cr
  & - F_k\partial \beta - F_k(\delta\partial \beta + \delta\beta_i\partial _0\beta_i) - 2a \delta\beta_i \partial _0\beta_i \bigg] dv^2 \cr
  & + 2r^2 \bigg[ (f-1)( \delta\beta_i + \frac12\delta^2\beta_i ) - a (\partial _0\beta_i + \delta\partial _0\beta_i + \delta\beta_j \partial _j\beta_i) - \frac{4r_H^3}{r^4}\delta r_H \delta\beta_i \cr
  & - F_k \partial \beta \delta\beta_i - F \delta\beta_j \partial _{(i}\beta_{j)} + \frac13 F \partial \beta\delta\beta_i \bigg] dvdx^i + 2\left(1 + \frac12 \delta\beta_i\delta\beta_i\right) dv dr \cr
  & + r^2 \bigg[ \left( 1 - \frac13 (F+\delta F) \partial \beta - \frac13 F (\delta\partial \beta + \delta\beta_k\partial _0\beta_k) \right) \delta_{ij} + (1-f)\delta\beta_i \delta\beta_j \cr
  & + 2a \delta\beta_{(i} \partial _{|0|}\beta_{j)} + (F+\delta F) \partial _{(i}\beta_{j)} + F (\delta\partial _{(i}\beta_{j)} + \delta\beta_{(i} \partial _{|0|}\beta_{j)}) \bigg] dx^idx^j \cr
  & - 2\left( \delta\beta_i + \frac12 \delta^2\beta_i \right) dx^i dr
\end{align}
The metric of the perturbative ansatz at 2nd order has the same form as (\ref{eq: 1st order perturbation ansatz}) with the first order perturbations changing to second order ones. By substituting (\ref{eq: 2nd order expanded metric D3}) and the 2nd order perturbations into the EOM (\ref{eq: EOM(ij)}) to (\ref{eq: EOM(rr)}) and the constraint equations \cref{eq: vector constraint,eq: scalar constraint 1,eq: scalar constraint 2}, we can get\footnote{Please note that there is no EOM for dilaton in D3-brane case.}
\begin{align}
  \partial_r\left( r^5f\partial_r\alpha_{ij}^{(2)} \right) =&\; (2r - 3r^2 F - 2r^3F') \left(\mathbf t_3 + \mathfrak T_1 + \frac{\mathfrak T_4}3 \right) - \left( \frac{4}{r^3} + 4r \right) \mathfrak T_5  \cr
  & + \left( 4r - 6 r^2 F + r^5 f F'^2 \right) \mathfrak T_6 - (4r + 6r^2F + 4r^3F') \mathfrak T_7, \label{eq: 2nd order EOM(ij) diff D3} \\
  \partial_r\left( r^5\partial_r w_{i}^{(2)} \right) =& -2r \mathbf v_4 + r^3 F' \mathbf v_5 + \left( \frac23r + \frac59r^3F' \right) \mathfrak V_1 + \left(  r - \frac56 r^3F' \right) \mathfrak V_2 \cr
  & - \left( r + \frac{11}{6} r^3F' \right) \mathfrak V_3, \label{eq: 2nd order EOM(ri) diff D3} \\
  rh''_{(2)} + 2h'_{(2)} - 2j'_{(2)} =&\; \frac{2}{3r^3} \mathfrak S_4 + \left( \frac16r F'^2 + \frac23FF' + \frac13rFF'' \right) \mathfrak S_5
  \label{eq: 2nd order EOM(rr) diff D3}
\end{align}
and
\begin{align}
  & \partial_i r_H^{(1)} = (2r^3 + r^5fF') \left( \frac14\mathbf v_5 + \frac{5}{36} \mathfrak V_1 - \frac{5}{24} \mathfrak V_2 - \frac{11}{24}\mathfrak V_3 \right), \label{eq: 2nd order vec constraint diff D3} \\
  & \partial_0 r_H^{(1)} = \frac{1}{3r} (\mathbf s_2 + \mathbf s_3 - \mathfrak S_1 - \mathfrak S_4) + \left( \frac{1}{3r} + \frac16 r^3 + \frac{1}{12} r^5f F' \right) \mathfrak S_5, \label{eq: 2nd order scalar constraint 1 diff D3} \\
  & (r^4k_{(2)})' + 8r^3j_{(2)} - (3r^4 - 1)h'_{(2)} = - \frac43r \mathbf s_2 + \frac23r \mathfrak S_1 - \frac29r \mathfrak S_3 + \left( \frac{1}{3r^3} + 2r \right) \mathfrak S_4 \cr
  & \hspace{6.26cm} - \left[ \frac53r + \frac13(3r^4-1) FF' + \frac{1}{12}r^5fF'^2 \right] \mathfrak S_5. \label{eq: 2nd order scalar constraint 2 diff D3}
\end{align}
During the procedure of casting the above differential equations, it needs to use the differential equations of first order perturbations like \cref{eq: 1st order EOM(ij) diff,eq: 1st order EOM(ri) diff,eq: 1st order EOM(rr) diff,eq: 1st order EOM(phi) diff,eq: 1st order scalar constraint 2 diff}, together with the Navier-Stokes equations \cref{eq: Navier-Stokes 0,eq: Navier-Stokes i} to eliminate the $x^\mu$ dependent terms.

Now we can use \cref{eq: integrate to get alpha&h,eq: integrate to get w} separately to get the solutions for the differential equations of tensor (\ref{eq: 2nd order EOM(ij) diff D3}) and vector (\ref{eq: 2nd order EOM(ri) diff D3}) perturbations:
\begin{align}
  \alpha_{ij}^{(2)} &= \frac{2-\ln2}{4 r^4} \left( \mathbf t_3 + \mathfrak T_1 + \frac{\mathfrak T_4}3 \right) + \frac{1}{r^2} \mathfrak T_5 + \left(  \frac{1}{r^2} + \frac{1}{2r^4} \right) \mathfrak T_6 + \left( \frac{2}{r^2} - \frac{\ln2}{2r^4} \right) \mathfrak T_7, \label{eq: 2nd order tensor perturbation solution D3} \\
  w_{i}^{(2)} &= \frac{1}{2r^2} \left( \mathbf v_4 + \mathbf v_5 \right) + \frac{1}{9r^2} \mathfrak V_1 - \frac{2}{3r^2} \left( \mathfrak V_2 + \mathfrak V_3 \right). \label{eq: 2nd order vector perturbation solution D3}
\end{align}
As for the scalar part, first we should use (\ref{eq: 2nd order EOM(rr) diff D3}) together with the gauge condition to get the differential equation for $h^{(2)}$, then solve it via
\begin{align}\label{eq: integrate to get h^(2) D3}
  h^{(2)} = \int_{r}^{\infty} \frac{1}{x^5} dx \int_{x}^{\infty} S_h(y) dy.
\end{align}
As one can see from the above that the way to get $h^{(2)}$ for D3-brane is the same as $w_i$ (\ref{eq: integrate to get w}), the other Dp-branes will follow the way of (\ref{eq: integrate to get alpha&h}). Then one can get $j^{(2)}$ and $k^{(2)}$ easily from (\ref{eq: integrate to get j&k}), the results of the scalar perturbations for $p=3$ are
\begin{align}\label{eq: 2nd order scalar perturbation solution D3}
  h ^{(2)} &= - \frac{1}{6r^2} \mathfrak S_4 - \frac{1}{6r^2} \mathfrak S_5, \cr
  j^{(2)} &= - \frac32 h ^{(2)} = \frac{1}{4r^2} \mathfrak S_4 + \frac{1}{4r^2} \mathfrak S_5, \cr
  k^{(2)} &= - \frac{2}{3r^2} \mathbf s_2 + \frac{1}{3r^2} \mathfrak S_1 - \frac{1}{9r^2} \mathfrak S_3 + \frac{1}{2r^2} \mathfrak S_4 + \frac{1}{2r^2} \mathfrak S_5.
\end{align}
We only retain the leading order in the large $r$ expansion in presenting the solutions of the vector and scalar perturbations. Because both of these two parts are trivial and only the leading order is relevant in calculating the stress-energy tensor of the dual fluid. The first scalar constraint (\ref{eq: 2nd order scalar constraint 1 diff D3}) and the vector constraint (\ref{eq: 2nd order vec constraint diff D3}) will reproduce the Navier-Stokes equations \cref{eq: Navier-Stokes 0,eq: Navier-Stokes i} after we expand $F(r)$.

\subsection{The D4-brane case}\label{subsec: solve D4-brane 2nd order perturbation}

Solving the D4-brane is almost the same like the compactified D4-brane case \cite{Wu1604}, we just need to keep in mind that the spatial dimension is 4 but not 3 for D4-brane case here. The differential equations for tensor perturbation is
\begin{align}\label{eq: D4 2nd diff eq tensor perturbation}
  \partial_r (r^4f\partial_r \alpha_{ij}^{(2)}) =& \left( 6r - \frac52 r^\frac32 F - 2r^\frac52 F' \right) \left( \mathbf t_3 + \mathfrak T_1 \right) + \bigg[ 2r + 2r^\frac32 F - \frac{21}5r^{\frac52} F'  \cr
  & - \frac45 r^\frac72 F'' + (4r^3-1)FF' + r^4fFF'' + 5r^\frac32F_j - 15r^2FF_j  \cr
  & + 2(4r^3-1)F'F_j - (5r^3-2) FF_j' + r^4f F'F_j' + 2r^4fF''F_j  \cr
  & - \frac{15}{2}r^2FF_k + 4r^3F'F_k - \frac{13}{2} r^3 FF_k' + r^4F'F_k' + r^4F''F_k - r^4 FF_k'' \bigg] \mathfrak T_4 \cr
  & - \left( \frac{4}{r^2} + 8r \right) \mathfrak T_5 + \left( 8r - 5r^\frac32F + r^4fF'^2 \right) \mathfrak T_6 \cr
  & - \left( 4r + 5r^\frac32F + 4r^\frac52F' \right) \mathfrak T_7,
\end{align}
Compared with the compactified D4-brane in \cite{Wu1604}, the above has the same coefficient functions for viscous tensors $\mathbf t_3,\mathfrak T_{1,5,6,7}$ (the definitions for $\mathfrak T_{5,6,7}$ are different from \cite{Wu1604}). Only the coefficient function of $\mathfrak T_4$ is different. Since coefficient functions determine the transport coefficients. Thus confirms that $\tau_\pi,\lambda_{1,2}$ of D4-brane are the same as the compactified D4-brane but $\tau_\pi^*$ is different.
The solution of (\ref{eq: D4 2nd diff eq tensor perturbation}) can be got from (\ref{eq: integrate to get alpha&h}) as
\begin{align}\label{eq: 2nd order tensor perturbation solution D4}
  \alpha_{ij}^{(2)} &= \left( \frac43 - \frac{\pi}{9\sqrt3} - \frac{\ln3}{3} \right) \frac{1}{r^3} \left( \mathbf t_3 + \mathfrak T_1 \right) + \left[ \frac2r + \left( \frac25 - \frac{\pi}{45\sqrt3} - \frac{\ln3}{15} \right) \frac{1}{r^3} \right] \mathfrak T_4 \cr
  & + \frac4r \mathfrak T_5 + \left( \frac4r + \frac{4}{3r^3} \right) \mathfrak T_6 + \left[ \frac8r - \left( \frac{2\pi}{9\sqrt3} + \frac{2\ln3}{3} \right) \frac{1}{r^3} \right] \mathfrak T_7.
\end{align}

The differential equation for the 2nd order vector perturbation of D4-brane is
\begin{align}
  \partial_r(r^4\partial_r w_i^{(2)}) =&\; \left( -2r -r^\frac52F' + \frac{10}{3}r^\frac32F_j - \frac43r^\frac52F_j' \right) \mathbf v_4 + \left( \frac{10}{3}r^\frac32F_j - \frac43r^\frac52F_j' \right) \mathbf v_5 \cr
  & + \left( \frac34r - \frac{29}{8}r^\frac52F' - 2r^\frac72F'' + \frac{45}{8}r^\frac32F_j + \frac{11}{4}r^\frac52F_j' - 2r^\frac72F_j'' \right) \mathfrak V_1 \cr
  & + \left( 5r - \frac32r^\frac52F' + \frac52r^\frac32F_j - r^\frac52F_j' \right) \mathfrak V_2 \cr
  & + \left( -r - \frac92r^\frac52F' + \frac52r^\frac32F_j - r^\frac52F_j' \right) \mathfrak V_3,
\end{align}
whose solution can be got from (\ref{eq: integrate to get w}) as
\begin{align}\label{eq: 2nd order vector perturbation solution D4}
  w_i^{(2)} = - \frac1r \mathfrak V_1 - \frac4r \mathfrak V_2 - \frac4r \mathfrak V_3 + \mathcal O\left( \frac{1}{r^{7/2}} \right).
\end{align}
Here we only keep the leading order for the vector perturbation, the same as D3-brane. The D1 and D2-brane will be treated similarly. The vector constraint is
\begin{align}\label{eq: 2nd order vector constraint diff D4}
  \partial _i r_H^{(1)} &= \left[ -\frac43r^\frac52 - \frac23r^4fF' + \frac89(5r^3-2)F_j + \frac{20}{9}r^3F_k + \frac89r^4F_k' \right]\mathbf v_4 \cr
  & + \left[ \frac89(5r^3-2)F_j + \frac{20}{9}r^3F_k + \frac89r^4F_k' \right] \mathbf v_5 \cr
  & + \left[ \frac{35}{6}r^\frac52 - \frac{11}{12}rF' + \frac{35}{12}r^4F' - \frac53F_j - \frac{95}{6} r^3 F_j - \frac{95}{12}r^3F_k - \frac52r^4F_k' \right]\mathfrak V_1 \cr
  & + \left[ -2r^\frac52 - r^4fF' + \frac23(5r^3-2)F_j + \frac53r^3F_k + \frac23r^4F_k' \right] \mathfrak V_2 \cr
  & + \left[ -6r^\frac52 - 3r^4fF' + \frac23(5r^3-2)F_j + \frac53r^3F_k + \frac23r^4F_k' \right] \mathfrak V_3.
\end{align}
After expanded, (\ref{eq: 2nd order vector constraint diff D4}) becomes
\begin{align}
  \partial_i r_H^{(1)} = \frac{4}{45} \mathbf v_4 + \frac{64}{45} \mathbf v_5 + \frac12 \mathfrak V_1 - \frac{14}{15} \mathfrak V_2 - \frac{74}{15} \mathfrak V_3,
\end{align}
which is just (\ref{eq: Navier-Stokes i}) with $p=4$.

In the scalar sector, the first scalar perturbation
\begin{align}\label{eq: 2nd order scalar perturbation 1 diff D4}
  \partial _0 r_H^{(1)} &= \left( \frac{4}{5r^\frac12} + \frac{4}{15}r^\frac52 - \frac15F + \frac{2}{15}r^4fF' - \frac23r^3fF_j - \frac13r^3F_k \right) \mathbf s_2 + \frac{2}{5r^\frac12} \mathbf s_3 \cr
  &  + \left( - \frac{8}{5r^\frac12} + \frac{4}{15}r^\frac52 - \frac15F + \frac{2}{15}r^4fF' - \frac23r^3fF_j - \frac13r^3F_k  \right) \mathfrak S_1 \cr
  & + \left( \frac{1}{25r^\frac12} + \frac{1}{15}r^\frac52 - \frac{1}{25}F + \frac{17}{150}r^4fF' + \frac{4}{75}r^5fF'' - \frac{2}{15}r^3fF_j - \frac{4}{15}r^4fF_j' \right. \cr
  & \left. - \frac{1}{15}r^3F_k \right) \mathfrak S_3 - \frac{4}{5r^\frac12} \mathfrak S_4 + \left( \frac{4}{5r^\frac12} + \frac{4}{15}r^\frac52 + \frac{2}{15}r^4fF' \right) \mathfrak S_5
\end{align}
will reproduce the scalar component of Navier-Stokes equations (\ref{eq: Navier-Stokes 0}) with $p=4$. To get this, one needs to expand $F,F_j$ and $F_k$ in (\ref{eq: 2nd order scalar perturbation 1 diff D4}) and has:
\begin{align}
  \partial _0 r_H^{(1)} &= \frac{1}{75} \mathfrak S_3 + \frac{4}{15} \mathfrak S_5 + \frac{4}{5r^{1/2}} \left( \mathbf s_2 + \frac12 \mathbf s_3 - 2\mathfrak S_1 + \frac{1}{20} \mathfrak S_3 - \mathfrak S_4 + \mathfrak S_5 \right).
\end{align}
The part inside the parenthesis is just the constraints (\ref{eq: 2nd order constraint (ii)}). Thus we reproduce (\ref{eq: Navier-Stokes 0}) for the case of $p=4$.

 In order to solve $h,j,k$ at the second order, we use the second scalar perturbation (\ref{eq: scalar constraint 2}):
\begin{align}
  & 5(r^3k_{(2)})' + 30r^2j_{(2)} - 4(5r^3-2)h_{(2)}' = \left( -16r + 5r^\frac32 F + 2r^\frac52F' \right) \mathbf s_2 \cr
  & + \left( 24r + 5r^\frac32F + 2r^\frac52F' \right) \mathfrak S_1 + \bigg[ - \frac52r + r^\frac52F + \frac{26}{5}r^\frac52F' + \frac45r^\frac72F'' + \frac38r^4fF'^2 \cr
  & - \frac14(5r^3-2)FF' - 10r^\frac32 F_j + 45r^2F_j^2 - 2(5r^3-2)F'F_j - 5r^3F'F_k - r^4F'F_k' \cr
  & + 30r^2F_jF_k + 10r^3F_jF_k' \bigg] \mathfrak S_3 + \left( \frac{2}{r^2} + 20r \right) \mathfrak S_4 \cr
  & - \left[ 18r + (5r^3-2)FF' + \frac12r^4fF'^2 \right] \mathfrak S_5,
\end{align}
the $(rr)$ component of Einstein equation (\ref{eq: EOM(rr)}):
\begin{align}
  4rh_{(2)}'' + 6h_{(2)}' - 5j_{(2)}' &= \left( \frac18rF'^2 + \frac38FF' + \frac14rFF'' - 5F_jF_j' + rF'F_j' \right) \mathfrak S_3 \cr
  & + \frac{2}{r^2} \mathfrak S_4 + \left( \frac12rF'^2 + \frac32FF' + rFF'' \right) \mathfrak S_5,
\end{align}
and the dilaton's EOM:
\begin{align}
  & ~\quad (r^3k_{(2)})' + r^3fj_{(2)}' + 6r^2j_{(2)} - 2r^3fh_{(2)}' = \left( -2r + \frac12r^\frac32F \right) \mathbf s_2 + \left( 6r + \frac12r^\frac32F \right) \mathfrak S_1   \cr
  & + \bigg[  - \frac12r + \frac{1}{10}r^\frac32F + \frac15r^\frac52F' - \frac18r^3fFF' - r^\frac32F_j + 9r^2F_j^2 + 3r^3fF_jF_j' - r^3fF'F_j  \cr
  & - \frac12r^3F'F_k + 6r^2F_jF_k + r^3F_j'F_k + 2r^3F_jF_k'  \bigg] \mathfrak S_3 + 2r\mathfrak S_4 - \left( 2r + \frac12r^3fFF' \right) \mathfrak S_5
\end{align}
to eliminate $j$ and $k$ and get the different equation for $h_{(2)}$, then $h_{(2)}$ can be solved by using (\ref{eq: integrate to get alpha&h}).
The final results for the 2nd order scalar perturbations of D4-brane are
\begin{align}\label{eq: 2nd order scalar perturbation solution D4}
  h ^{(2)} =& \left( \frac13 - \frac{\pi}{36\sqrt3} - \frac{\ln3}{12} \right) \frac{1}{r^3} ( \mathbf s_2 + \mathfrak S_1 ) + \left[ \frac{1}{4r} + \left( \frac{1}{60} - \frac{\pi}{180\sqrt3} - \frac{\ln3}{60} \right)\frac{1}{r^3} \right] \mathfrak S_3 \cr
  & + \frac1r \mathfrak S_4 + \left( \frac1r + \frac{1}{3r^3} \right) \mathfrak S_5, \cr
  j^{(2)} =& - \left( \frac23 - \frac{\pi}{18\sqrt3} - \frac{\ln3}{6} \right) \frac{1}{r^3} ( \mathbf s_2 + \mathfrak S_1 ) - \left( \frac{1}{30} - \frac{\pi}{90\sqrt3} - \frac{\ln3}{30} \right) \frac{1}{r^3} \mathfrak S_3 \cr
  & - \frac{2}{3r^3} \mathfrak S_5, \cr
  r^3 k^{(2)} =& \left[ - \left( \frac{4}{15} - \frac{\pi}{45\sqrt3} - \frac{\ln3}{15} \right) + \frac{1}{10r} - \left( \frac23 - \frac{\pi}{18\sqrt3} -
  \frac{\ln3}{6} \right)\frac{1}{r^3} \right] \mathbf s_2 \cr
  & + \left[ 4r^2 - \left( \frac{4}{15} - \frac{\pi}{45\sqrt3} - \frac{\ln3}{15} \right) + \frac{1}{10r} - \left( \frac23 - \frac{\pi}{18\sqrt3} -
  \frac{\ln3}{6} \right)\frac{1}{r^3} \right] \mathfrak S_1 \cr
  & + \left[ - \left( \frac{1}{75} - \frac{\pi}{225\sqrt3} - \frac{\ln3}{75} \right) + \frac{2}{35r^{1/2}} + \frac{163}{100r} + \left( \frac{1}{90} +
  \frac{\pi}{90\sqrt3} + \frac{\ln3}{30} \right) \frac{1}{r^3} \right] \mathfrak S_3 \cr
  & - \frac2r \mathfrak S_4 - \left( \frac{4}{15} - \frac{8}{7r^{1/2}} - \frac1r + \frac{2}{3r^3} \right) \mathfrak S_5.
\end{align}
Here the constants at the right hand side of the solution of $k^{(2)}$ are integration constants, which are fixed by requiring the boundary stress-energy tensor is in Landau frame. We do not need to fix the integration constants for D3-brane because the scalar perturbations are trivial and do not contribute to the stress-energy tensor.

\subsection{The D1-brane case}\label{subsec: D1 2nd order}

The D1-brane case is very different from the other cases for no viscous tensors can exist in this case. We list the viscous scalars and vectors for $p=1$ in \cref{tab: D1 2nd order viscous terms list}. Note we define a new viscous vector $\mathbf v_3 = \partial_1^2\beta_1$ for D1-brane out of a linear combination of $\mathbf v_4$ and $\mathbf v_5$ since both of them can not exist for $p=1$. The viscous terms for D1-brane case satisfy the following 4 constraints:
\begin{align}
  &\mathbf s_1 + \frac14 \mathbf s_2 - \frac14 \mathfrak S_1 - \frac{1}{16} \mathfrak S_3 = 0, \label{eq: 2nd constraint (00) D1} \\
  &\mathbf s_2 + 2 \mathbf s_3 - \frac12 \mathfrak S_1 + \frac12 \mathfrak S_3 = 0, \label{eq: 2nd constraint (ii) D1} \\
  &\mathbf v_1 + \frac12 \mathbf v_2 + \frac18 \mathfrak V_1  = 0, \label{eq: 2nd constraint (0i) D1} \\
  &\mathbf v_1 + \frac14 \mathbf v_3 - \frac38 \mathfrak V_1 = 0, \label{eq: 2nd constraint (i0) D1}
\end{align}
which can be got from the general form of the constraints \cref{eq: 2nd order constraint (00),eq: 2nd order constraint (0i),eq: 2nd order constraint (i0),eq: 2nd order constraint (ii)} by specifying $p=1$.

\begin{table}
\centering
\begin{tabular}{|l|l|l|}
  \hline
  Scalars of $\mathrm{SO}(1)$ & Vectors of $\mathrm{SO}(1)$  \\
  \hline\hline
  $\mathbf{s}_1 = \frac1{r_H} \partial_0^2 r_H$ & $\mathbf{v}_{1} = \frac1{r_H} \partial_0\partial_1 r_H$  \\
  $\mathbf{s}_2 = \partial_0\partial_1\beta_1$ & $\mathbf{v}_{2} = \partial_0^2\beta_1$  \\
  $\mathbf{s}_3 = \frac1{r_H} \partial_1^2 r_H$ & $\mathbf{v}_3 = \partial_1^2\beta_1$  \\
  $\mathfrak S_1 = \partial_0\beta_1\partial_0\beta_1$ & $\mathfrak V_{1} = \partial_0\beta_1\partial_1\beta_1$  \\
  $\mathfrak S_3 = (\partial_1\beta_1)^2$ &   \\
  \hline
\end{tabular}
\caption{\label{tab: D1 2nd order viscous terms list} The list of all the second order viscous terms for D1-brane. The vector viscous terms are reduced to scalars in this case but we still call them vectors.}
\end{table}

There is no tensor perturbation for D1-brane, thus we start from the vector perturbation. The vector constraint for D1-brane is
\begin{align}
  \partial _1 r_H^{(1)} &= \left[ - \frac16F_j + \frac23r^6F_j + \frac13r^6F_k + \frac1{12}r^7F_k' \right] \mathbf v_3 \cr
  & + \left[ \frac16r^4 + \frac1{24}rF' + \frac1{12}r^7F' - \frac13(2r^6+1)F_j - \frac13r^6F_k \right] \mathfrak V_1.
\end{align}
After expansion the above becomes
\begin{align}
  \partial _1 r_H^{(1)} = \frac{1}{12} \mathbf v_3,
\end{align}
which is just the spatial component of the Navier-Stokes equation (\ref{eq: Navier-Stokes i}). The vector dynamical equation is
\begin{align}
  \partial _r(r^7 \partial _r w_1^{(2)}) = \left( 4r^3F_j - r^4F_j' \right) \mathbf v_3 + \left( - \frac52r^4F' - \frac12r^5F'' + 12r^3F_j + \frac12r^4F_j' - \frac12r^5F_j'' \right) \mathfrak V_1
\end{align}
from which we get
\begin{align}
  w_1^{(2)} = - \frac{1}{4r^4} \mathfrak V_1.
\end{align}

The scalar constraint for D1-brane is
\begin{align}
  \partial _0 r_H^{(1)} &= \left( \frac{1}{8r^2} + \frac1{12}r^4 - \frac18F + \frac1{24}r^7fF' - \frac13r^6fF_j - \frac16r^6F_k \right) \mathbf s_2 + \frac{1}{4r^2} \mathbf s_3 \cr
  &  + \left( - \frac{1}{16r^2} + \frac1{12}r^4 - \frac18F + \frac1{24}r^7fF' - \frac13r^6fF_j - \frac16r^6F_k \right) \mathfrak S_1 \cr
  & + \left( \frac{1}{16r^2} + \frac1{12}r^4 - \frac{1}{16}F + \frac7{96}r^7fF' + \frac1{96}r^8fF'' - \frac16r^6fF_j - \frac1{12}r^7fF_j' \right. \cr
  & \left. - \frac{1}{12}r^6F_k \right) \mathfrak S_3
\end{align}
under the large $r$ expansion one has
\begin{align}
  \partial _0 r_H^{(1)} = \frac{1}{24} \mathfrak S_3 + \frac{1}{8r^2} \left( \mathbf s_2 + 2\mathbf s_3 - \frac12 \mathfrak S_1 + \frac12 \mathfrak S_3 \right)
\end{align}
The terms inside the parenthesis in the above equation is just the 2nd order constraint of D1-brane (\ref{eq: 2nd constraint (ii) D1}) thus gives no contribution. So we get the scalar component of the Navier-Stokes equation for D1-brane as
\begin{align}
  \partial _0 r_H^{(1)} = \frac{1}{24} \mathfrak S_3,
\end{align}
which can be got from (\ref{eq: Navier-Stokes 0}) by setting $p=1$ and remove $\mathfrak S_5$ (there is no shear tensor for D1-brane). The dynamical equation for the scalar perturbations are the second scalar constraint (\ref{eq: scalar constraint 2})
\begin{align}
  & ~~\quad 4(r^6k_{(2)})' + 48r^5j_{(2)} - (4r^6-1)h_{(2)}' \cr
  & = \left( -2r + 4r^3F + r^4F' \right) \mathbf s_2 + \left( 4r^3F + r^4F' \right) \mathfrak S_1 + \bigg[ - 2r + 2r^3F + \frac{11}4r^4F' + \frac14r^5F'' \cr
  & - (4r^6-1)FF' - 8r^3F_j + 72r^5F_j^2 - 2(4r^6-1)F'F_j - 4r^6F'F_k - \frac12r^7F'F_k' \cr
  & + 48r^5F_jF_k + 8r^6F_jF_k' \bigg] \mathfrak S_3,
\end{align}
the $(rr)$ component of Einstein equation (\ref{eq: EOM(rr)})
\begin{align}
  rh_{(2)}'' + 3h_{(2)}' - 8j_{(2)}' &= \left( \frac12rF'^2 + 3FF' + rFF'' - 8F_jF_j' + rF'F_j' \right) \mathfrak S_3,
\end{align}
and the EOM of $\phi$ (\ref{eq: EOM(phi)})
\begin{align}
  & ~\quad 2(r^6k_{(2)})' + 2r^6f j_{(2)}' + 24r^5 j_{(2)} - r^6f h_{(2)}' = \left( -r + r^3F \right) \mathbf s_2 + r^3F \mathfrak S_1  \cr
  & + \bigg[ - r + \frac12r^3F + \frac14r^4F' - r^6fFF' - 2r^3F_j + 36r^5F_j^2 + 6r^6fF_jF_j' - 2r^6fF'F_j  \cr
  & - r^6F'F_k + 24r^5F_jF_k + 2r^6F_j'F_k + 4r^6F_jF_k' \bigg] \mathfrak S_3,
\end{align}
from which we can solve all the scalar perturbations as
\begin{align}
  h^{(2)} =& \left( \frac16 + \frac{\pi}{36\sqrt3} - \frac{\ln3}{12} \right) \frac{1}{r^6} (\mathbf s_2 + \mathfrak S_1) + \left[ \frac{1}{4r^4} + \left( \frac{1}{12} + \frac{\pi}{72\sqrt3} - \frac{\ln3}{24} \right) \frac{1}{r^6} \right] \mathfrak S_3, \\
  j^{(2)} =& - \left( \frac{1}{12} + \frac{\pi}{72\sqrt3} - \frac{\ln3}{24} \right) \frac{1}{r^6} (\mathbf s_2 + \mathfrak S_1) - \left( \frac{1}{24} + \frac{\pi}{144\sqrt3} - \frac{\ln3}{48} \right) \frac{1}{r^6} \mathfrak S_3, \\
  r^6 k^{(2)} =& - \left[ \left( \frac{1}{12} + \frac{\pi}{72\sqrt3} - \frac{\ln3}{24} \right) - \frac{1}{40r^4} + \left( \frac{1}{12} + \frac{\pi}{72\sqrt3} - \frac{\ln3}{24} \right) \frac{1}{r^6} \right] \mathbf s_2 \cr
  & + \left[ \frac14 r^2 - \left( \frac{1}{12} + \frac{\pi}{72\sqrt3} - \frac{\ln3}{24} \right) + \frac{1}{40r^4} - \left( \frac{1}{12} + \frac{\pi}{72\sqrt3} - \frac{\ln3}{24} \right) \frac{1}{r^6} \right] \mathfrak S_1 \cr
  & - \left[ \left( \frac{1}{24} + \frac{\pi}{144\sqrt3} - \frac{\ln3}{48} \right) - \frac{1}{16r^2} - \frac{1}{16r^4} \right. \cr
  & \left. + \left( \frac{1}{72} + \frac{\pi}{144\sqrt3} - \frac{\ln3}{48} \right) \frac{1}{r^6} \right] \mathfrak S_3.
\end{align}

\subsection{The D2-brane case}\label{subsec: D2 2nd order}

The D2-brane case is the most complicated one and that's why we leave it to the last. The differential equation for the tensor perturbation of D2-brane is
\begin{align}
  \partial_r \left( r^6f \partial_r\alpha_{ij}^{(2)} \right) =& \left( \frac23r - \frac72r^\frac52F - 2r^\frac72F' \right) \left( t_3 + \mathfrak T_1 \right) + \bigg[ \frac23r - \frac32r^\frac52F - \frac{31}7r^\frac72F' \cr
  & - \frac47r^\frac92F'' + \frac12r^6fF'^2 + (6r^5-1)FF' + r^6fFF'' + 7r^\frac52F_j - 35r^4FF_j  \cr
  & + 2(6r^5-1)F'F_j + 2r^6fF''F_j - (7r^5-2)FF_j' + r^6f F'F_j'  - \frac{35}{2}r^4FF_k \cr
  & + 6r^5F'F_k + r^6F''F_k - \frac{19}{2}r^5FF_k' + r^6F'F_k' - r^6FF_k'' \bigg] \mathfrak T_4 \cr
  & - \left( 4r + 7r^\frac52F + 4r^\frac72F' \right) \mathfrak T_7.
\end{align}
Note in the source term there is no $\mathfrak T_{5,6}$ at present since both of them are zero for $p=2$. Using the relations between $F_{j,k}$ and $F$ (\ref{eq: Dp 1st order hjk solutions}) and the differential equation for $F$ (\ref{eq: 1st order EOM(ij) diff}) at the case $p=2$ will simplify the source term and make it easy to be integrated. The solution can be got from (\ref{eq: integrate to get alpha&h}) as
\begin{align}
  \alpha_{ij}^{(2)}(r) &= \left( \frac{4}{15} + \frac{\pi}{25} \sqrt{1 - \frac2{\sqrt5}} + \frac{1}{5\sqrt5} \,{\rm arcoth}\,\sqrt5 - \frac{\ln5}{10} \right) \frac{1}{r^5} (\mathbf t_3 + \mathfrak T_1) \cr
  & + \left[ \frac{4}{9r^3} + \left( \frac{16}{105} + \frac{3\pi}{175} \sqrt{1 - \frac2{\sqrt5}} + \frac{3}{35\sqrt5} \,{\rm arcoth}\,\sqrt5 - \frac{3}{70} \ln5 \right) \frac{1}{r^5} \right] \mathfrak T_4 \cr
  & + \left[ \frac{8}{9r^3} + \left( \frac{2\pi}{25} \sqrt{1 - \frac2{\sqrt5}} + \frac{2}{5\sqrt5} \,{\rm arcoth}\,\sqrt5 - \frac{\ln5}{5} \right) \frac{1}{r^5} \right] \mathfrak T_7.
\end{align}
As one can see from the above that the solution of D2-brane is indeed more complicated than the other nonconformal situations: except the terms proportional to $\pi$ and logarithm, there appears another arcoth term.

The 2nd order vector constraint for D2-brane is
\begin{align}
  \partial _i r_H^{(1)} &= \left[ -\frac4{15} r^\frac72 - \frac2{15} r^6 f F' + \frac4{15} (7r^5-2) F_j + \frac{14}{15} r^5 F_k + \frac4{15} r^6 F_k' \right] \mathbf v_4 \cr
  & + \left[ \frac4{15} (7r^5-2) F_j + \frac{14}{15} r^5 F_k + \frac4{15} r^6 F_k' \right] \mathbf v_5 \cr
  & + \left[ \frac{23}{45} r^\frac72 - \frac1{30} r F' + \frac{23}{90} r^6 F' - \frac25 F_j - \frac{77}{45} r^5 F_j - \frac{77}{90} r^5 F_k - \frac19 r^6 F_k' \right] \mathfrak V_1 \cr
  & + \left[ - \frac25r^\frac72 - \frac15 r^6 fF' + \frac2{15} (7r^5-2) F_j + \frac7{15} r^5 F_k + \frac2{15} r^6 F_k' \right] \mathfrak V_2 \cr
  & + \left[ - \frac{22}{45} r^\frac72 - \frac{11}{45} r^6 f F' + \frac2{15} (7r^5-2) F_j + \frac7{15} r^5 F_k + \frac2{15} r^6 F_k' \right] \mathfrak V_3.
\end{align}
After large $r$ expansion, the above becomes
\begin{align}
  \partial _i r_H^{(1)} = \frac{4}{105} \mathbf v_4 + \frac{32}{105} \mathbf v_5 + \frac{37}{315} \mathfrak V_1 - \frac{26}{105} \mathfrak V_2 - \frac{106}{315} \mathfrak V_3.
\end{align}
This is just the vector component of Navier-Stokes equation (\ref{eq: Navier-Stokes i}) when $p=2$. The vector dynamical equation is
\begin{align}
  \partial _r(r^6\partial _r w_i^{(2)}) &= \left( - 2r - r^\frac72F' + 7r^\frac52F_j - 2r^\frac72F_j' \right) \mathbf v_4 + \left( 7r^\frac52F_j - 2r^\frac72F_j' \right) \mathbf v_5 \cr
  & + \left( \frac12r - \frac{25}{12}r^\frac72F' - \frac23r^\frac92F'' + \frac{35}{4}r^\frac52F_j + \frac76r^\frac72F_j' - \frac23r^\frac92F_j'' \right) \mathfrak V_1 \cr
  & + \left( - \frac13r - \frac32r^\frac72F' + \frac72r^\frac52F_j - r^\frac72F_j' \right) \mathfrak V_2 \cr
  & + \left( - r - \frac{11}6r^\frac72F' + \frac72r^\frac52F_j - r^\frac72F_j' \right) \mathfrak V_3,
\end{align}
whose leading order solution is
\begin{align}
  w_i^{(2)} = - \frac{2}{9r^3} \mathfrak V_1 - \frac{4}{9r^3} \mathfrak V_2 - \frac{4}{9r^3} \mathfrak V_3.
\end{align}

The first scalar constraint for $p=2$ is
\begin{align}
  \partial _0 r_H^{(1)} &= \left( \frac4{21 r^{3/2}} + \frac4{35} r^\frac72 - \frac17 F + \frac2{35} r^6f F' - \frac25 r^5f F_j - \frac15 r^5 F_k \right) \mathbf s_2 + \frac{2}{7r^{3/2}} \mathbf s_3 \cr
  &  + \left( - \frac8{63r^{3/2}} + \frac4{35} r^\frac72 - \frac17 F + \frac2{35} r^6f F' - \frac25 r^5f F_j - \frac15 r^5 F_k \right) \mathfrak S_1 \cr
  & + \left( \frac2{147 r^{3/2}} + \frac{2}{35} r^\frac72 - \frac3{49} F + \frac{17}{245} r^6f F' + \frac{4}{245} r^7f F'' - \frac6{35} r^5f F_j - \frac4{35}r^6 f F_j' \right. \cr
  &\left. - \frac3{35} r^5 F_k \right) \mathfrak S_3 - \frac4{21r^{3/2}} \mathfrak S_4 + \left( \frac4{21r^{3/2}} + \frac4{35} r^\frac72 + \frac2{35} r^6 f F' \right) \mathfrak S_5.
\end{align}
After expansion, the above will reproduce the scalar component of the Navier-Stokes equation (\ref{eq: Navier-Stokes 0}) for $p=2$:
\begin{align}
  \partial _0 r_H^{(1)} = \frac{2}{245} \mathfrak S_3 + \frac{4}{35} \mathfrak S_5 + \frac{4}{21r^\frac32} \left( \mathbf s_2 + \frac32 \mathbf s_3 - \frac23 \mathfrak S_1 + \frac{1}{14} \mathfrak S_3 - \mathfrak S_4 + \mathfrak S_5 \right).
\end{align}
The terms in the parenthesis of the above is just the second order constraint (\ref{eq: 2nd order constraint (ii)}) at $p=2$ thus gives 0. So we finally have
\begin{align}
  \partial _0 r_H^{(1)} = \frac{2}{245} \mathfrak S_3 + \frac{4}{35} \mathfrak S_5.
\end{align}
The above equation is just the scalar component of Navier-Stokes equation at $p=2$.

The scalar dynamical equations are still the second scalar constraint (\ref{eq: scalar constraint 2})
\begin{align}
  & ~\quad 7(r^5k_{(2)})' + 70r^4j_{(2)} - 2(7r^5-2)h_{(2)}' \cr
  & = \left( -\frac{16}3r + 7r^\frac52F + 2r^\frac72F' \right) \mathbf s_2 + \left( \frac89r + 7r^\frac52F + 2r^\frac72F' \right) \mathfrak S_1 + \bigg[ - \frac53r + 3r^\frac52F \cr
  & + \frac{38}7r^\frac72F' + \frac47r^\frac92F'' + \frac14r^6fF'^2 - \frac12(7r^5-2)FF' - 14r^\frac52F_j + 105r^4F_j^2 \cr
  & - 2(7r^5-2)F'F_j - 7r^5F'F_k - r^6F'F_k' + 70r^4F_jF_k + 14r^5F_jF_k' \bigg] \mathfrak S_3 \cr
  & + \left( \frac2{r^4} + \frac{28}{3}r \right) \mathfrak S_4 - \left[ \frac{22}3r + \frac12r^6fF'^2 + (7r^5-2)FF' \right] \mathfrak S_5,
\end{align}
the $(rr)$ component of the Einstein equation (\ref{eq: EOM(rr)})
\begin{align}
  2r h_{(2)}'' + 5h_{(2)}' - 7j_{(2)}' &= \left( \frac14rF'^2 + \frac54FF' + \frac12rFF'' - 7F_jF_j' + rF'F_j' \right) \mathfrak S_3 \cr
  & + \frac{2}{r^4} \mathfrak S_4 + \left( \frac12rF'^2 + \frac52FF' + rFF'' \right) \mathfrak S_5,
\end{align}
and the EOM of dilaton (\ref{eq: EOM(phi)})
\begin{align}
  & ~\quad (r^5k_{(2)})' + r^5f j_{(2)}' + 10r^4 j_{(2)} - r^5f h_{(2)}' = \left( -\frac23r + \frac12r^\frac52F \right) \mathbf s_2  \cr
  & + \left( \frac29r + \frac12r^\frac52F \right) \mathfrak S_1 + \bigg[  - \frac13r + \frac{3}{14}r^\frac52F + \frac17r^\frac72F' - \frac14r^5fFF' - r^\frac52F_j + 15r^4F_j^2  \cr
  & - r^5fF'F_j + 3r^5fF_jF_j' - \frac12r^5F'F_k + 10r^4F_jF_k + r^5F_j'F_k + 2r^5F_jF_k'  \bigg] \mathfrak S_3 \cr
  & + \frac23r \mathfrak S_4 - \left( \frac23r + \frac12r^5fFF' \right) \mathfrak S_5.
\end{align}
From which we can solve the scalar perturbations for $p=2$ as
\begin{align}
  h^{(2)} &= \left( \frac{2}{15} + \frac{\pi}{50} \sqrt{1 - \frac2{\sqrt5}} + \frac{1}{10\sqrt5} \,{\rm arcoth}\,\sqrt5 - \frac{\ln5}{20} \right) \frac{1}{r^5} (\mathbf s_2 + \mathfrak S_1) \cr
  & + \left[ \frac{1}{9r^3} + \left( \frac{1}{105} + \frac{3\pi}{350} \sqrt{1 - \frac2{\sqrt5}} + \frac{3}{70\sqrt5} \,{\rm arcoth}\,\sqrt5 - \frac{3}{140} \ln5 \right) \frac{1}{r^5} \right] \mathfrak S_3 \cr
  & + \frac{2}{9r^3} \mathfrak S_4 + \left( \frac{2}{9r^3} + \frac{2}{15r^5} \right) \mathfrak S_5, \\
  j^{(2)} &= - \left( \frac{2}{15} + \frac{\pi}{50} \sqrt{1 - \frac2{\sqrt5}} + \frac{1}{10\sqrt5} \,{\rm arcoth}\,\sqrt5 - \frac{\ln5}{20} \right) \frac{1}{r^5} (\mathbf s_2 + \mathfrak S_1) \cr
  & - \left( \frac{1}{105} + \frac{3\pi}{350} \sqrt{1 - \frac2{\sqrt5}} + \frac{3}{70\sqrt5} \,{\rm arcoth}\,\sqrt5 - \frac{\ln5}{140} \right) \frac{1}{r^5} \mathfrak S_3 - \frac{2}{15r^5} \mathfrak S_5, \\
  r^5 k^{(2)} &= \bigg[ - \left( \frac{4}{35} + \frac{3\pi}{175} \sqrt{1 - \frac2{\sqrt5}} + \frac{3}{35\sqrt5} \,{\rm arcoth}\,\sqrt5 - \frac{3}{70} \ln5 \right) + \frac{1}{28r^3} \cr
  & - \left( \frac{2}{15} + \frac{\pi}{50} \sqrt{1 - \frac2{\sqrt5}} + \frac{1}{10\sqrt5} \,{\rm arcoth}\,\sqrt5 - \frac{\ln5}{20} \right) \frac{1}{r^5} \bigg] \mathbf s_2 \cr
  & + \bigg[ \frac49 r^2 - \left( \frac{4}{35} + \frac{3\pi}{175} \sqrt{1 - \frac2{\sqrt5}} + \frac{3}{35\sqrt5} \,{\rm arcoth}\,\sqrt5 - \frac{3}{70} \ln5 \right) + \frac{1}{28r^3} \cr
  & - \left( \frac{2}{15} + \frac{\pi}{50} \sqrt{1 - \frac2{\sqrt5}} + \frac{1}{10\sqrt5} \,{\rm arcoth}\,\sqrt5 - \frac{\ln5}{20} \right) \frac{1}{r^5} \bigg] \mathfrak S_1 \cr
  & + \bigg[ - \left( \frac{2}{245} + \frac{9\pi}{1225} \sqrt{1 - \frac2{\sqrt5}} + \frac{9}{245\sqrt5} \,{\rm arcoth}\,\sqrt5 - \frac{9}{490} \ln5 \right) + \frac{4}{273r^\frac32} \cr
  & + \frac{17}{147r^3} + \left( \frac{13}{525} - \frac{3\pi}{350} \sqrt{1 - \frac2{\sqrt5}} - \frac{3}{70\sqrt5} \,{\rm arcoth}\,\sqrt5 + \frac{3}{140} \ln5 \right) \frac{1}{r^5} \bigg] \mathfrak S_3 \cr
  & - \frac{2}{9r^3} \mathfrak S_4 + \left( - \frac{4}{35} + \frac{8}{39r^\frac32} + \frac{1}{18r^3} - \frac{2}{15r^5} \right) \mathfrak S_5.
\end{align}

\section{The second order stress-energy tensor}\label{sec: 2nd order stress-energy tensor}

\subsection{Formulation of the result}\label{subsec: Dp 2nd order stress-energy tensor}

Now, using the second order expanded metric (\ref{eq: 2nd order expanded metric Dp}) and (\ref{eq: 2nd order expanded metric D3}) together with the already solved second order perturbations of Dp-brane, we can calculate the boundary stress-energy tensor via (\ref{eq: Dp-brane Brown-York tensor}). We will only offer the results here, the readers who are interested in the details can refer to \cite{Wu1604}.

In order to get the covariant form of the stress-energy tensor, we will need the following rules to substitute the 2nd order spatial viscous tensors:
\begin{align}
  &\mathbf t_{3ij}=\partial_0\sigma_{ij} \to \sideset{_\langle}{}{\mathop D}\partial_\mu u_{\nu\rangle},~~\mathfrak T_{1ij} = \partial_0\beta_i \partial_0\beta_j - \frac1p \delta_{ij} \mathfrak S_1 \to Du_{\langle\mu}Du_{\nu\rangle}, \cr
  &\mathfrak T_{4ij} = \sigma_{ij}\partial\beta \to \sigma_{\mu\nu} \partial_\rho u^\rho,~~\mathfrak T_{6ij} = \sigma_{ik}\sigma_{kj} - \frac1p \delta_{ij} \mathfrak S_5 \to \sigma_{\langle\mu}^{~~\rho} \sigma_{\nu\rangle\rho}, \cr
  & \mathfrak T_{7ij} = \sigma_{(i}^{~~k} \Omega_{j)k} \to \sigma_{\langle\mu}^{~~\rho} \Omega_{\nu\rangle\rho}
\end{align}
Then the boundary stress-energy tensor for D3-brane under full consideration of dimension can be calculated as
\begin{align}\label{eq: 2nd order stress-energy tensor D3}
  T_{\mu\nu} &= \frac1{2\kappa_5^2} \Bigg\{ \frac{r_H^4}{L_3^5} ( 3u_\mu u_\nu + P_{\mu\nu} ) - \left( \frac{r_H}{L_3} \right)^3 \cdot 2\sigma_{\mu\nu} \cr
  & + \frac{r_H^2}{L_3} \bigg[ \frac{2-\ln2}2 \cdot 2\bigg( \sideset{_\langle}{}{\mathop D} \sigma_{\mu\nu\rangle} + \frac13 \sigma_{\mu\nu} \partial u \bigg) + \frac12 \cdot 4\sigma_{\langle\mu}^{~~\rho}\sigma_{\nu\rangle\rho} - \ln2 \cdot 2\sigma_{\langle\mu}^{~~\rho}
  \Omega_{\nu\rangle\rho} \bigg] \Bigg\}. \quad
\end{align}
Compare the above with (\ref{eq: relativistic fluid 2nd general stress tensor}) one can get the 2nd order transport coefficients for D3-brane
\begin{align}\label{eq: 2nd order transport coefficients D3}
  \eta\tau_\pi = \frac1{2\kappa_5^2} \frac{2-\ln2}2 \frac{r_H^2}{L_3},~~~~\lambda_1 = \frac1{2\kappa_5^2} \frac12 \frac{r_H^2}{L_3},~~~~\lambda_2 = - \frac1{2\kappa_5^2} \ln2 \frac{r_H^2}{L_3}.
\end{align}
We have restored the parameters $r_H,~L_p$ and $\kappa_{p+2}^2$ which has length dimension in \cref{eq: 2nd order stress-energy tensor D3,eq: 2nd order transport coefficients D3}. The results in (\ref{eq: 2nd order transport coefficients D3}) have already been derived in \textit{e.g.} \cite{Bhattacharyya0712,Baier0712,Arnold1105}.

As for the D4-brane, we need further substitution rules for the scalar viscous terms
\begin{align}
  \mathbf s_2 + \mathfrak S_1 = \partial_0 \partial \beta + \partial_0\beta_i \partial_0\beta_i \to D(\partial u),~~\mathfrak S_3 = (\partial \beta)^2 \to (\partial u)^2,~~\mathfrak S_5 = \sigma_{ij}^2 \to \sigma_{\mu\nu}^2.
\end{align}
The boundary stress-energy tensor for D4-brane can then be formulated as
\begin{align}\label{eq: D4 2nd order stress-energy tensor}
  T_{\mu\nu} &= \frac1{2\kappa_6^2} \Bigg\{ \frac{r_H^3}{L_4^4} \bigg( \frac52 u_\mu u_\nu + \frac12 P_{\mu\nu} \bigg) - \left( \frac{r_H}{L_4} \right)^\frac52 \bigg( 2\sigma_{\mu\nu} + \frac1{10} \partial_\rho u^\rho P_{\mu\nu} \bigg) \cr
  & + \frac{r_H^2}{L_4} \bigg[ \bigg( 2 - \frac\pi{6\sqrt3} - \frac{\ln3}2 \bigg)\cdot 2\bigg( \sideset{_\langle}{}{\mathop D}\sigma_{\mu\nu\rangle} + \frac14
  \sigma_{\mu\nu} \partial u \bigg) + \bigg( \frac25+\frac{\pi}{30\sqrt3} + \frac{\ln3}{10} \bigg) \frac{ 2\sigma_{\mu\nu}\partial u }{4} \cr
  & + 4\sigma_{\langle\mu}^{~~\rho}\sigma_{\nu\rangle\rho} - \bigg( \frac{\pi}{3\sqrt3} + \ln3\bigg ) \cdot 2\sigma_{\langle\mu}^{~~\rho}
  \Omega_{\nu\rangle\rho} \bigg] + \frac{r_H^2}{L_4} P_{\mu\nu} \bigg[ \bigg( \frac15 - \frac{\pi}{60\sqrt3} - \frac{\ln3}{20} \bigg) D(\partial u) \cr
  & + \bigg( \frac1{100} - \frac{\pi}{300\sqrt3} - \frac{\ln3}{100} \bigg) (\partial u)^2 + \frac1{20}\cdot 4\sigma_{\alpha\beta}^2 \bigg] \Bigg\},
\end{align}
from which we can read all the second order transport coefficients as
\begin{align}\label{eq: D4 2nd order transport coefficients}
  \eta\tau_\pi &= \frac1{2\kappa_6^2} \bigg( 2 - \frac\pi{6\sqrt3} - \frac{\ln3}2 \bigg) \frac{r_H^2}{L_4}, ~\eta\tau_\pi^* = \frac1{2\kappa_6^2} \bigg( \frac25 + \frac{\pi}{30\sqrt3} + \frac{\ln3}{10} \bigg) \frac{r_H^2}{L_4}, ~ \lambda_1 = \frac1{2\kappa_6^2} \frac{r_H^2}{L_4}, \quad \cr
  \lambda_2 &= - \frac1{2\kappa_6^2} \bigg( \frac{\pi}{3\sqrt3} + \ln3\bigg ) \frac{r_H^2}{L_4}, ~ \zeta\tau_\Pi = \frac1{2\kappa_6^2} \bigg( \frac15 - \frac{\pi}{60\sqrt3} - \frac{\ln3}{20} \bigg) \frac{r_H^2}{L_4}, ~ \xi_1 = \frac1{2\kappa_6^2} \frac{1}{20} \frac{r_H^2}{L_4}, \cr
  \xi_2 &= \frac1{2\kappa_6^2} \bigg( \frac1{100} - \frac{\pi}{300\sqrt3} - \frac{\ln3}{100} \bigg) \frac{r_H^2}{L_4}.
\end{align}
Compared with the results for compactified D4-brane \cite{Wu1604}, one can see that $\eta\tau_\pi$ and $\lambda_{1,2}$ which are the transport coefficients of viscous tensors do indeed not change, just as what we have discussed in \cref{subsec: Dp 1st order,subsec: solve D4-brane 2nd order perturbation}. But the other 4 coefficients \textit{i.e.}, $\eta\tau_\pi^*$, $\zeta\tau_\Pi$ and $\xi_{1,2}$ which associate with nonconformal viscous terms are different from that in \cite{Wu1604}. This may lie in the difference of spatial dimensions. Further more, $\eta\tau_\pi^*$ and $\xi_{1}$ are separately half and $\frac38$ of that for compactified D4-brane, while for $\zeta\tau_\Pi$ and $\xi_{2}$ there are no simple ratios.

As one can check, the shear and bulk relaxation time of D4-brane (\ref{eq: D4 2nd order transport coefficients}) that we get here satisfies the relation in equation (4.7) of \cite{Natsuume0712} after using the Hawking temperature (\ref{eq: Hawking temperature of Dp}) with $p=4$.

The relativistic fluid dual to D1-brane has the stress-energy tensor
\begin{align}\label{eq: D1 2nd order stress-energy tensor}
  T_{\mu\nu} &= \frac1{2\kappa_3^2} \Bigg\{ \frac{r_H^6}{L_1^7} (4u_\mu u_\nu + 2P_{\mu\nu}) - \left( \frac{r_H}{L_1} \right)^4 P_{\mu\nu} \partial u \cr
  & + \frac{r_H^2}{L_1} P_{\mu\nu} \bigg[ \left( \frac12 + \frac{\pi}{12\sqrt3} - \frac{\ln3}{4} \right) D(\partial u) + \left( \frac14 + \frac{\pi}{24\sqrt3} - \frac{\ln3}{8} \right) (\partial u)^2 \bigg] \Bigg\}.
\end{align}
It has only two 2nd order transport coefficients from the scalar viscous terms
\begin{align}\label{eq: D1 2nd transport coefficients}
  \zeta \tau_\Pi = \frac1{2\kappa_3^2} \left( \frac12 + \frac{\pi}{12\sqrt3} - \frac{\ln3}{4} \right) \frac{r_H^2}{L_1},~~\xi_2 = \frac1{2\kappa_3^2} \left( \frac14 + \frac{\pi}{24\sqrt3} - \frac{\ln3}{8} \right) \frac{r_H^2}{L_1}.
\end{align}
One can check that our results of the 2nd order transport coefficients of D1-brane (\ref{eq: D1 2nd transport coefficients}) agree with equation (4.6) of \cite{Natsuume0712} where the author derived the bulk relaxation time $\tau_\Pi$ for D1-brane.

The stress-energy tensor of D2-brane is
\begin{align}\label{eq: D2 2nd order stress-energy tensor}
  T_{\mu\nu} &= \frac{1}{2\kappa_{4}^{2}} \left\{ \frac{r_H^5}{L_2^6} \left( \frac72 u_\mu u_\nu + \frac32 P_{\mu\nu} \right) - \left( \frac{r_H}{L_2} \right)^\frac72 \left( 2 \sigma_{\mu\nu} + \frac17 P_{\mu\nu} \partial u \right) \right. \cr
  & + \frac{r_H^2}{L_2} \left[ \left( \frac23 + \frac{\pi}{10} \sqrt{1 - \frac2{\sqrt5}} + \frac{1}{2\sqrt5} \,{\rm arcoth}\,\sqrt5 - \frac{\ln5}{4} \right) 2 \left( \sideset{_\langle}{}{\mathop D} \sigma_{\mu\nu\rangle} + \frac12 \sigma_{\mu\nu} \partial_\rho u^\rho \right) \right. \cr
  & + \left( \frac{2}{21} - \frac{\pi}{70} \sqrt{1 - \frac2{\sqrt5}} - \frac{1}{14\sqrt5} \,{\rm arcoth}\,\sqrt5 + \frac{\ln5}{28} \right) 2 \cdot \frac{\sigma_{\mu\nu} \partial_\rho u^\rho}{2} \cr
  & + \left. \left( \frac{\pi}{5} \sqrt{1 - \frac2{\sqrt5}} + \frac{1}{\sqrt5} \,{\rm arcoth}\,\sqrt5 - \frac{\ln5}{2} \right) 2\sigma_{\langle\mu}^{~~\rho} \Omega_{\nu\rangle\rho} \right] \cr
  & + \frac{r_H^2}{L_2} P_{\mu\nu} \left[ \left( \frac{2}{21} + \frac{\pi}{70} \sqrt{1 - \frac2{\sqrt5}} + \frac{1}{14\sqrt5} \,{\rm arcoth}\,\sqrt5 - \frac{\ln5}{28} \right) D\partial u \right. \cr
  & + \left. \left. \left( \frac{1}{147} + \frac{3\pi}{490} \sqrt{1 - \frac2{\sqrt5}} + \frac{3}{98\sqrt5} \,{\rm arcoth}\,\sqrt5 - \frac{3 \ln5}{196} \right) (\partial u)^2 + \frac{1}{42} \cdot 4\sigma_{\alpha\beta}^2 \right] \right\}. \qquad
\end{align}
The 2nd order transport coefficients can be read from the above as
\begin{align}\label{eq: D2 2nd order coefficients}
  \eta\tau_\pi &= \frac1{2\kappa_4^2} \left( \frac23 + \frac{\pi}{10} \sqrt{1 - \frac2{\sqrt5}} + \frac{1}{2\sqrt5} \,{\rm arcoth}\,\sqrt5 - \frac{\ln5}{4} \right) \frac{r_H^2}{L_2}, \cr
  \eta\tau_\pi^* &= \frac1{2\kappa_4^2} \left( \frac{2}{21} - \frac{\pi}{70} \sqrt{1 - \frac2{\sqrt5}} - \frac{1}{14\sqrt5} \,{\rm arcoth}\,\sqrt5 + \frac{\ln5}{28} \right) \frac{r_H^2}{L_2}, \cr
  \lambda_2 &= \frac1{2\kappa_4^2} \left( \frac{\pi}{5} \sqrt{1 - \frac2{\sqrt5}} + \frac{1}{\sqrt5} \,{\rm arcoth}\,\sqrt5 - \frac{\ln5}{2} \right) \frac{r_H^2}{L_2}, \cr
  \zeta\tau_\Pi &= \frac1{2\kappa_4^2} \left( \frac{2}{21} + \frac{\pi}{70} \sqrt{1 - \frac2{\sqrt5}} + \frac{1}{14\sqrt5} \,{\rm arcoth}\,\sqrt5 - \frac{\ln5}{28} \right) \frac{r_H^2}{L_2},~~\xi_1 = \frac1{2\kappa_4^2} \frac{1}{42} \frac{r_H^2}{L_2}, \cr
  \xi_2 &= \frac1{2\kappa_4^2} \left( \frac{1}{147} + \frac{3\pi}{490} \sqrt{1 - \frac2{\sqrt5}} + \frac{3}{98\sqrt5} \,{\rm arcoth}\,\sqrt5 - \frac{3}{196} \ln5 \right) \frac{r_H^2}{L_2}.
\end{align}
Compared with the 2nd order coefficients of D4-brane, the D2-brane does not have $\lambda_1$. This is due to the fact that $\sigma_{\langle\mu}^{~~\rho}\sigma_{\nu\rangle\rho}$ can not exist in 3d relativistic hydrodynamics. In the literature, \cite{Natsuume0807} also gives the explicit result of the shear relaxation time of D2-brane (as in equation (4.6) of \cite{Natsuume0807}). But by comparing, we find that the shear relaxation time of D2-brane of our results (\ref{eq: D2 2nd order coefficients}) does not agree with that in \cite{Natsuume0807}. We think that the result offered in \cite{Natsuume0807} should be wrong which will be specified later.

The 2nd order stress-energy tensor for Dp-brane with $1\leq p \leq 4$ can be nicely written in the following unified form in terms of Harmonic number:
\begin{align}\label{eq: Dp 2nd order stress-energy tensor (general)}
  T_{\mu\nu} &= \frac{1}{2 \kappa_{p+2}^2} \Bigg\{ {r_H^{7-p} \over L_p^{8-p}} \left( \frac{9-p}{2} u_\mu u_\nu + \frac{5-p}{2} P_{\mu\nu} \right) - \left( \frac{r_H}{L_p} \right) ^\frac{9-p}{2} \bigg( 2\sigma_{\mu\nu} + \frac{2(p-3)^2}{p(9-p)} P_{\mu\nu} \partial u \bigg) \cr
  & + \frac{r_H^2}{L_p} \Bigg[ \bigg( \frac{1}{5-p} + \frac{1}{7-p} H_\frac{5-p}{7-p} \bigg)\cdot 2\bigg( \sideset{_\langle}{}{\mathop D}\sigma_{\mu\nu\rangle} + \frac1p \sigma_{\mu\nu} \partial u \bigg) \cr
  & + \bigg( \frac{3(p-3)^2}{(5-p)(9-p)} - \frac{(p-3)^2}{(7-p)(9-p)} H_\frac{5-p}{7-p} \bigg) \frac{2\sigma_{\mu\nu} \partial u}{p} + \frac{1}{5-p} \cdot 4\sigma_{\langle\mu}^{~~\rho}\sigma_{\nu\rangle\rho} \cr
  & + \bigg( - \frac{2}{5-p} + \frac{2}{7-p} H_\frac{5-p}{7-p} \bigg) \cdot 2 \sigma_{\langle\mu}^{~~\rho} \Omega_{\nu\rangle\rho} \Bigg] \cr
  & + \frac{r_H^2}{L_p} P_{\mu\nu} \Bigg[ \bigg( \frac{2(p-3)^2}{p(5-p)(9-p)} + \frac{2(p-3)^2}{p(7-p)(9-p)} H_\frac{5-p}{7-p} \bigg) D(\partial u) \cr
  & + \bigg( \frac{2(p-3)^2(3p^2 - 17p + 18)}{p^2 (5-p)(9-p)^2} + \frac{2(p-3)^2 (5-p)}{p(7-p)(9-p)^2} H_\frac{5-p}{7-p} \bigg) (\partial u)^2 \cr
  & + \frac{(p-3)^2}{p(5-p)(9-p)} \cdot 4 \sigma_{\alpha\beta}^2 \Bigg]
  \Bigg\}.
\end{align}
Note the surface gravity in the above results for Dp-brane is $\kappa_{p+2}^2$ which has the mass dimension $-p$. This is consistent with the mass dimension of the stress-energy tensor of Dp-brane \textit{i.e.} $p+1$. Then one can read the general form of all the 2nd order transport coefficients as
\begin{align}\label{eq: Dp 2nd order coefficients}
  \eta\tau_\pi &= \frac1{2\kappa_{p+2}^2} \left( \frac{1}{5-p} + \frac{1}{7-p} H_\frac{5-p}{7-p} \right) \frac{r_H^2}{L_p}, \cr
  \eta\tau_\pi^* &= \frac1{2\kappa_{p+2}^2} \left( \frac{3(p-3)^2}{(5-p)(9-p)} - \frac{(p-3)^2}{(7-p)(9-p)} H_\frac{5-p}{7-p} \right) \frac{r_H^2}{L_p}, \cr
  \lambda_1 &= \frac1{2\kappa_{p+2}^2} \frac{1}{5-p} \frac{r_H^2}{L_p},~~\lambda_2 = \frac1{2\kappa_{p+2}^2} \left( - \frac{2}{5-p}
  + \frac{2}{7-p} H_\frac{5-p}{7-p} \right) \frac{r_H^2}{L_p}, \cr
  \zeta\tau_\Pi &= \frac1{2\kappa_{p+2}^2} \left( \frac{2(p-3)^2}{p(5-p)(9-p)} + \frac{2(p-3)^2}{p(7-p)(9-p)} H_\frac{5-p}{7-p} \right) \frac{r_H^2}{L_p}, \cr
  \xi_1 &= \frac1{2\kappa_{p+2}^2} \frac{(p-3)^2}{p(5-p)(9-p)} \frac{r_H^2}{L_p}, \cr
  \xi_2 &= \frac1{2\kappa_{p+2}^2} \left( \frac{2(p-3)^2(3p^2 - 17p + 18)}{p^2 (5-p)(9-p)^2} + \frac{2(p-3)^2 (5-p)}{p(7-p)(9-p)^2} H_\frac{5-p}{7-p} \right) \frac{r_H^2}{L_p}.
\end{align}
Here $H_\frac{5-p}{7-p}$ is the Harmonic number which is defined as
\begin{align}
  H_\frac ab = \frac ba + 2 \sum_{n=1}^{\left[ \frac{b-1}{2} \right]} \cos\left( \frac{2\pi n a}{b} \right) \ln\sin\left( \frac{n\pi}{b} \right) - \frac{\pi}{2} \cot\left( \frac{\pi a}{b} \right) - \ln(2b),
\end{align}
where $a,b$ are positive integers with $a<b$. For the special cases of $1\leq p\leq 4$, one has
\begin{align}
  H_{\frac23} &= \frac32 + \frac{\pi}{2\sqrt3} - \frac32 \ln3, \quad (p=1) \\
  H_{\frac35} &= \frac53 + \frac\pi2 \sqrt{1 - \frac2{\sqrt5}} + \frac{5}{2\sqrt5} \,{\rm arcoth}\, \sqrt5 - \frac54 \ln5, \quad (p=2) \\
  H_{\frac12} &= 2 - 2 \ln2, \quad (p=3) \\
  H_{\frac13} &= 3 - \frac{\pi}{2\sqrt3} - \frac32 \ln3. \quad (p=4)
\end{align}
Please remember that $\eta\tau_\pi,\eta\tau_\pi^*,\lambda_2,\xi_1$ are not appropriate for $p=1$ and $\lambda_1$ is not suitable for both $p=1,2$ in (\ref{eq: Dp 2nd order coefficients}).

We find previous studies \cite{Springer0902} and \cite{Natsuume0807} also give general expressions on Dp-brane relaxation time. The author of \cite{Springer0902} calculates the sound mode dispersion relation for Dp-branes and finds a relation between the shear and bulk relaxation time as in its equation (79). One can check that our results (\ref{eq: Dp 2nd order coefficients}) satisfy that relation. In equation (4.5) of \cite{Natsuume0807}, the author offers a general formula for the shear relaxation time of Dp-brane. If one compares the result of $\tau_\pi$ in (\ref{eq: Dp 2nd order coefficients}) with that from \cite{Natsuume0807}, one can see that the result in \cite{Natsuume0807} misses a term of $\frac{1}{5-p}$. That is how $\tau_\pi$ of D2-brane in \cite{Natsuume0807} disagrees with our result in (\ref{eq: D2 2nd order coefficients}).

The way to express all the results for Dp-brane \textit{i.e.} the energy density, pressure and the first order transport coefficients in (\ref{eq: Dp 1st order transport coefficients}), as well as the second order transport coefficients listed in (\ref{eq: Dp 2nd order coefficients}) are all expressed via geometric parameters $2\kappa_{p+2}^2$, $r_H$ and $L_p$. One can also reformulate the expressions in field theory language via 't Hooft coupling of the gauge theory on Dp-brane world-volume $\lambda_{p+1}$, the number of Dp-branes $N$ and the temperature of the boundary fluid $T$. $\lambda_{p+1}$ here is defined as $\lambda_{p+1}= g_{p+1}^2 N \Lambda^{p-3}$, where $g_{p+1}^2= (2\pi)^{p-2} g_s l_s ^{p-3}$ is the gauge coupling of the effective field theory on Dp-brane world-volume and $\Lambda= r/\alpha'$ is the characteristic UV cutoff scale under the field theory limit. With the definitions of $\lambda_{p+1}$, $g_{p+1}^2$, $L_p$ (\ref{eq: definition of L_p}) and Hawking temperature (\ref{eq: Hawking temperature of Dp}), we can translate the geometric language of expressing the results into field theory language. The results are listed in \cref{tab: all the results in field theory language}.

\begin{table}
\renewcommand\arraystretch{1.75}
\centering
\begin{tabular}{|c|c|c|}
\hline
 $\varepsilon$     & $\frac{9-p}{2} \left( \frac{4\pi}{7-p} \right)^2 { 2^\frac{2(7-p)}{5-p} \pi^\frac{3-p}{5-p} \over (7-p)^\frac{11-p}{5-p} } \Gamma\left( \frac{9-p}2 \right)^\frac{2}{5-p} \lambda_{p+1}^\frac{p-3}{5-p} N^2 {T^\frac{2(7-p)}{5-p} \over \Lambda^\frac{(p-3)^2}{5-p}}$ \\[2mm] 
\hline
 $\mathfrak p$    & $\frac{5-p}{2} \left( \frac{4\pi}{7-p} \right)^2 { 2^\frac{2(7-p)}{5-p} \pi^\frac{3-p}{5-p} \over (7-p)^\frac{11-p}{5-p} } \Gamma\left( \frac{9-p}2 \right)^\frac{2}{5-p} \lambda_{p+1}^\frac{p-3}{5-p} N^2 {T^\frac{2(7-p)}{5-p} \over \Lambda^\frac{(p-3)^2}{5-p}}$ \\[2mm]
\hline
 $\eta$                & $\left( \frac{4\pi}{7-p} \right) { 2^\frac{2(7-p)}{5-p} \pi^\frac{3-p}{5-p} \over (7-p)^\frac{11-p}{5-p} } \Gamma\left( \frac{9-p}2 \right)^\frac{2}{5-p} \lambda_{p+1}^\frac{p-3}{5-p} N^2 {T^\frac{9-p}{5-p} \over \Lambda^\frac{(p-3)^2}{5-p}}$ \\[2mm]
\hline
 $\zeta$              & $\frac{2(p-3)^2}{p(9-p)} \left( \frac{4\pi}{7-p} \right) { 2^\frac{2(7-p)}{5-p} \pi^\frac{3-p}{5-p} \over (7-p)^\frac{11-p}{5-p} } \Gamma\left( \frac{9-p}2 \right)^\frac{2}{5-p} \lambda_{p+1}^\frac{p-3}{5-p} N^2 {T^\frac{9-p}{5-p} \over \Lambda^\frac{(p-3)^2}{5-p}}$ \\[2mm]
\hline
 $\eta\tau_\pi$   & $\left( \frac{1}{5-p} + \frac{1}{7-p} H_\frac{5-p}{7-p} \right) { 2^\frac{2(7-p)}{5-p} \pi^\frac{3-p}{5-p} \over (7-p)^\frac{11-p}{5-p} } \Gamma\left( \frac{9-p}2 \right)^\frac{2}{5-p} \lambda_{p+1}^\frac{p-3}{5-p} N^2 {T^\frac4{5-p} \over \Lambda^\frac{(p-3)^2}{5-p}}$ \\[2mm]
\hline
 $\eta\tau_\pi^*$ & $\left( \frac{3(p-3)^2}{(5-p)(9-p)} - \frac{(p-3)^2}{(7-p)(9-p)} H_\frac{5-p}{7-p} \right) { 2^\frac{2(7-p)}{5-p} \pi^\frac{3-p}{5-p} \over (7-p)^\frac{11-p}{5-p} } \Gamma\left( \frac{9-p}2 \right)^\frac{2}{5-p} \lambda_{p+1}^\frac{p-3}{5-p} N^2 {T^\frac4{5-p} \over \Lambda^\frac{(p-3)^2}{5-p}}$ \\ [2mm]
\hline
 $\lambda_1$     & $\frac{1}{5-p} { 2^\frac{2(7-p)}{5-p} \pi^\frac{3-p}{5-p} \over (7-p)^\frac{11-p}{5-p} } \Gamma\left( \frac{9-p}2 \right)^\frac{2}{5-p} \lambda_{p+1}^\frac{p-3}{5-p} N^2 {T^\frac4{5-p} \over \Lambda^\frac{(p-3)^2}{5-p}}$ \\ [2mm]
\hline
 $\lambda_2$     & $\left( - \frac{2}{5-p} + \frac{2}{7-p} H_\frac{5-p}{7-p} \right) { 2^\frac{2(7-p)}{5-p} \pi^\frac{3-p}{5-p} \over (7-p)^\frac{11-p}{5-p} } \Gamma\left( \frac{9-p}2 \right)^\frac{2}{5-p} \lambda_{p+1}^\frac{p-3}{5-p} N^2 {T^\frac4{5-p} \over \Lambda^\frac{(p-3)^2}{5-p}}$ \\ [2mm]
\hline
 $\zeta\tau_\Pi$  & $\left( \frac{2(p-3)^2}{p(5-p)(9-p)} + \frac{2(p-3)^2}{p(7-p)(9-p)} H_\frac{5-p}{7-p} \right) { 2^\frac{2(7-p)}{5-p} \pi^\frac{3-p}{5-p} \over (7-p)^\frac{11-p}{5-p} } \Gamma\left( \frac{9-p}2 \right)^\frac{2}{5-p} \lambda_{p+1}^\frac{p-3}{5-p} N^2 {T^\frac4{5-p} \over \Lambda^\frac{(p-3)^2}{5-p}}$ \\ [2mm]
\hline
 $\xi_1$     & $\frac{(p-3)^2}{p(5-p)(9-p)} { 2^\frac{2(7-p)}{5-p} \pi^\frac{3-p}{5-p} \over (7-p)^\frac{11-p}{5-p} } \Gamma\left( \frac{9-p}2 \right)^\frac{2}{5-p} \lambda_{p+1}^\frac{p-3}{5-p} N^2 {T^\frac4{5-p} \over \Lambda^\frac{(p-3)^2}{5-p}}$ \\ [2mm]
\hline
 $\xi_2$     & $\left( \frac{2(p-3)^2(3p^2 - 17p + 18)}{p^2 (5-p)(9-p)^2} + \frac{2(p-3)^2 (5-p)}{p(7-p)(9-p)^2} H_\frac{5-p}{7-p} \right) { 2^\frac{2(7-p)}{5-p} \pi^\frac{3-p}{5-p} \over (7-p)^\frac{11-p}{5-p} } \Gamma\left( \frac{9-p}2 \right)^\frac{2}{5-p} \lambda_{p+1}^\frac{p-3}{5-p} N^2 {T^\frac4{5-p} \over \Lambda^\frac{(p-3)^2}{5-p}}$ \\ [2mm]
\hline
\end{tabular}
\caption{\label{tab: all the results in field theory language} Reformulation of the results in field theory language. Remember that $\lambda_3$ and $\xi_3$ are both 0.}
\end{table}

\subsection{Relations between second order transport coefficients}\label{subsec: relations of 2nd order coefficients}

Now we would like to discuss the identities satisfied by the 2nd order transport coefficients calculated in this paper.

First is the Haack-Yarom (HY) relation $4\lambda_1 + \lambda_2 = 2\eta\tau_\pi$. As it has been discussed in \cite{Wu1604} that the HY relation has been verified to be valid for a wide range of cases except for the second order $\lambda_{GB}$ correction \cite{Shaverin1509,Grozdanov2015,Grozdanov1611}. By the 2nd order stress-energy tensor of D1 (\ref{eq: D1 2nd order stress-energy tensor}) and D2-brane (\ref{eq: D2 2nd order stress-energy tensor}) we can see that the HY relation does not exist for these two cases. Because D1-brane does not support any tensor viscous term and D2-brane does not have the term $\sigma_{\langle\mu}^{~~\rho}\sigma_{\nu\rangle\rho}$ which relates $\lambda_1$. Similar observation has also been made in \cite{Haack0811} for $\mathrm U(1)$ charged conformal fluid. But the HY relation holds for D3 and D4-brane which can be seen from their 2nd order transport coefficients \textit{i.e.} \cref{eq: 2nd order transport coefficients D3,eq: D4 2nd order transport coefficients}. Thus we can conclude that the HY relation exists for Dp-brane with only $p>2$ and is also satisfied by those cases. If one does not consider D5 and D6-brane that do not have dual relativistic fluid, we find D4-brane as a new example that confirms the robustness of HY relation in nonconformal regime.

The next are two relations that were first proposed by Romatschke in \cite{Romatschke0906}:
\begin{align}
  \tau_\pi &= \tau_\Pi, \label{eq: tau_pi = tau_Pi} \\
  \xi_1 &= \frac{1-3c_s^2}3 \lambda_1. \label{eq: relation between xi_1 and lambda_1}
\end{align}
It has been shown that the above are satisfied by the 2nd order transport coefficients of compactified D4-brane in \cite{Wu1604}. What we want to point out first is (\ref{eq: tau_pi = tau_Pi}) is not suitable for D1-brane and (\ref{eq: relation between xi_1 and lambda_1}) is only valid for $p>2$. Then if we check the above two relations using (\ref{eq: Dp 2nd order coefficients}), we can find (\ref{eq: tau_pi = tau_Pi}) still holds while (\ref{eq: relation between xi_1 and lambda_1}) is not. This is because Romatschke only considered the hydrodynamics of 4d in \cite{Romatschke0906}. The second relation has to be generalized to arbitrary dimensions as
\begin{align}\label{eq: relation between xi_1 and lambda_1 (generalized)}
  \xi_1 = \frac{1}{d-1} \left( 1 - (d-1) c_s^2 \right) \lambda_1, \quad (d>3)
\end{align}
where $d=p+1$ is the dimension of the relativistic fluid. As one can check that (\ref{eq: Dp 2nd order coefficients}) satisfies (\ref{eq: relation between xi_1 and lambda_1 (generalized)}).

Besides the two relations discussed in the last paragraph, \cite{Romatschke0906} also proposed many other relations between the 2nd order coefficients, \textit{e.g.}
\begin{align}\label{eq: wrong relations in Romatschke0906}
  \tau_\pi^*=-(1-3c_s^2)\tau_\pi, \quad \xi_2=\frac{1-3c_s^2}{3}2c_s^2\eta\tau_\pi.
\end{align}
But it has been pointed out in \cite{Kleinert1610} by Kleinert and Probst that these two relations may miss the term of $\lambda_1$, and the correct form should be
\begin{align}\label{eq: Kleinert-Probst relation}
  \eta\tau^*_\pi = (1-3c_s^2) (4\lambda_1-\eta\tau_\pi),~~~~\xi_2 = \frac29 (1-3c_s^2) [3c_s^2 \eta\tau_\pi + (1-6c_s^2)2\lambda_1].
\end{align}
Kleinert and Probst have also checked that the above two relations are satisfied by the compactified D4-brane \cite{Benincasa0605,Wu1508,Wu1604}. But if we are in dimensions other than 4 like the case in this paper, the above Kleinert-Probst (KP) relations (\ref{eq: Kleinert-Probst relation}) should be generalized to
\begin{align}\label{eq: Kleinert-Probst relation (generalized)}
  \eta\tau^*_\pi &= \left(1 - (d-1) c_s^2 \right) (4\lambda_1 - \eta\tau_\pi), \quad (d>3) \cr
  \xi_2 &= \frac2{(d-1)^2} \left(1 - (d-1) c_s^2 \right) \left[ (1 - 2(d-1) c_s^2) 2\lambda_1 + (d-1) c_s^2 \eta\tau_\pi \right]. ~ (d>3) \quad
\end{align}
As one can check, the above two generalized KP relations are indeed satisfied by the 2nd order transport coefficients of Dp-brane (\ref{eq: Dp 2nd order coefficients}) in this paper. Since the above two relations also have $\lambda_1$, so they are only appropriate for the cases with $p>2$ (or $d>3$), like the HY relation and the generalized Romatschke relation (\ref{eq: relation between xi_1 and lambda_1 (generalized)}).

There are some other nontrivial relations among the 2nd order thermodynamical transport coefficients\footnote{As it has been classified in \cite{Moore1210} that there are 8 thermodynamical transport coefficients out of the 15 second order coefficients which do not cause the increase of entropy.} proposed in \cite{Romatschke0906}, which are
\begin{align}\label{eq: Romatschke relation curved}
  \kappa^* &= - \frac{1 - 3 c_s^2}{2 c_s^2} \kappa, \qquad \xi_3 = \frac{1 - 3 c_s^2}{3} \lambda_3, \cr
   \xi_5 &= \frac{1 - 3 c_s^2}{3} \kappa, \qquad \xi_6 = \frac{1 - 3 c_s^2}{3 c_s^2} \kappa.
\end{align}
Kleinert and Probst check the above relations to be true for the 5d Chamblin-Reall background using the method of \cite{Kanitscheider0901}. According to the generalization from (\ref{eq: Kleinert-Probst relation}) to (\ref{eq: Kleinert-Probst relation (generalized)}), we think the above relations may also need to be generalized in arbitrary dimensions to:
\begin{align}\label{eq: Romatschke relation curved (generalized)}
  \kappa^* &= - \frac{1 - (d-1) c_s^2}{2 c_s^2} \kappa, \qquad \xi_3 = \frac{1 - (d-1) c_s^2}{d-1} \lambda_3, \cr
   \xi_5 &= \frac{1 - (d-1) c_s^2}{d-1} \kappa, \qquad \xi_6 = \frac{1 - (d-1) c_s^2}{(d-1) c_s^2} \kappa.
\end{align}
Although we do not know for now whether (\ref{eq: Romatschke relation curved (generalized)}) is completely correct since it needs to be checked by strict calculation in nonconformal backgrounds. We hope future studies will give us a satisfied answer.

\subsection{Test of Kanitscheider-Skenderis proposal}\label{subsec: test of KS proposal}

Kanitscheider and Skenderis proposed a way of generating the second order part of nonconformal stress-energy tensor from that of a higher dimensional conformal relativistic fluid \cite{Kanitscheider0901}. With the help of the results in this paper, we can test whether their method is correct.

First we need the result of stress-energy tensor of conformal relativistic fluid in various dimensions. The following stress-energy tensor summarizes all the results from \cite{Policastro0205,Policastro0210,Baier0712,Bhattacharyya0712,Barnes1004,Arnold1105,Herzog0210,Natsuume0712,Natsuume0801,
VanRaamsdonk0802,Haack0806,Bhattacharyya0809} and offers the stress-energy tensor of $d$-dimensional conformal relativistic fluid that corresponds to AdS$_{d+1}$ black hole:
\begin{align}\label{eq: AdS_(d+1) stress-energy tensor}
  T_{\mu\nu} &= \frac{1}{2\kappa_{d+1}^2} \left\{ \frac{r_H^d}{L^{d+1}} \left[ (d-1) u_\mu u_\nu + P_{\mu\nu} \right] - \left( \frac{r_H}{L} \right)^{d-1} 2\sigma_{\mu\nu} \right. \cr
  & + \frac{r_H^{d-2}}{L^{d-3}} \left[ \left( \frac12 + \frac{1}{d} H_{\frac{2}{d}} \right) 2\left( \sideset{_\langle}{}{\mathop D} \sigma_{\mu \nu \rangle} + \frac1{d-1} \sigma_{\mu\nu} \partial u \right) + \frac12 \cdot 4\sigma_{\langle \mu}^{~~\rho}\sigma_{\nu \rangle \rho} \right. \cr
  & + \left. \left. \left( \frac2d H_{\frac2d} - 1 \right) 2\sigma_{\langle \mu}^{~~\rho}\Omega_{\nu \rangle \rho} + \frac{2}{d-2} \left( R_{\langle \mu\nu \rangle} - 2u^\rho u^\sigma R_{\rho \langle \mu\nu \rangle \sigma} \right) \right] \right\},
\end{align}
where $\frac{1}{2\kappa_{d+1}^2}$, $L$ and $r_H$ are separately the surface gravity constant, the curvature scale and the location of horizon of AdS$_{d+1}$ black hole background. The energy density, pressure and the transport coefficients can then be read as\footnote{The result in \cite{Bhattacharyya0809} is expressed via $H_{\frac2d-1}$, which can be related to $H_{\frac{2}{d}}$ by the recurrence relation of Harmonic number: $H_\alpha = H_{\alpha-1} + \frac1\alpha$.}
\begin{align}
  \varepsilon &= \frac{1}{2\kappa_{d+1}^2} (d-1) \frac{r_H^d}{L^{d+1}}, \quad \mathfrak p = \frac{1}{2\kappa_{d+1}^2} \frac{r_H^d}{L^{d+1}}, \quad \eta = \frac{1}{2\kappa_{d+1}^2} \frac{r_H^{d-1}}{L^{d-1}},  \cr
  \eta\tau_\pi &= \frac{1}{2\kappa_{d+1}^2} \left( \frac12 + \frac1d H_{\frac2d} \right) \frac{r_H^{d-2}}{L^{d-3}}, \quad \kappa = \frac{1}{2\kappa_{d+1}^2} \frac{2}{d-2} \frac{r_H^{d-2}}{L^{d-3}}, \cr
   \lambda_1 &= \frac{1}{2\kappa_{d+1}^2} \frac12 \frac{r_H^{d-2}}{L^{d-3}}, \quad \lambda_2 = \frac{1}{2\kappa_{d+1}^2} \left( \frac2d H_{\frac2d} - 1 \right) \frac{r_H^{d-2}}{L^{d-3}}, \quad \lambda_3 = 0.
\end{align}
The Hawking temperature, entropy density and shear to entropy ratio are
\begin{align}
  T = \frac{d}{4\pi} \frac{r_H}{L^2}, \quad s = \frac{1}{2\kappa_{d+1}^2} 4\pi \frac{r_H^{d-1}}{L^{d-1}}, \quad \frac{\eta}{s} = \frac{1}{4\pi}.
\end{align}

Following the method that proposed by Kanitscheider and Skenderis in \cite{Kanitscheider0901}, the 2nd order viscous terms in (\ref{eq: AdS_(d+1) stress-energy tensor}) will transform like
\begin{align}\label{eq: Kanitscheider-Skenderis transformation}
  \sideset{_\langle}{}{\mathop D} \sigma_{\mu \nu \rangle} &\to \sideset{_\langle}{}{\mathop D} \sigma_{\mu \nu \rangle} + \chi P_{\mu\nu} D(\partial u), \quad \sigma_{\mu\nu} \partial u \to \sigma_{\mu\nu} \partial u + \chi P_{\mu\nu} (\partial u)^2, \cr
  \sigma_{\langle \mu}^{~~\rho}\sigma_{\nu \rangle \rho} &\to \sigma_{\langle \mu}^{~~\rho}\sigma_{\nu \rangle \rho} + \chi P_{\mu\nu} \sigma_{\alpha \beta}^2, \quad \sigma_{\langle \mu}^{~~\rho}\Omega_{\nu \rangle \rho} \to \sigma_{\langle \mu}^{~~\rho}\Omega_{\nu \rangle \rho}.
\end{align}
Here $\chi$ is defined as
\begin{align}
  \chi = \frac{\tilde d - d}{(\tilde d - 1)(d-1)}.
\end{align}
We denote $\tilde d$ as the dimension of conformal fluid that is dual to AdS$_{\tilde d + 1}$ black hole background, and $d=p+1$ as the dimension of the nonconformal fluid which correspond to Dp-brane, with $\tilde d >d$. Now we want to see with the transformation rule (\ref{eq: Kanitscheider-Skenderis transformation}), whether we can get the 2nd order part of stress-energy tensor for nonconformal $d$-dimensional fluid from that of a $\tilde d$-dimensional conformal fluid. The 2nd order part stress-energy tensor of $\tilde d$-dimensional conformal fluid is
\begin{align}
  \widetilde{\eta\tau_\pi} \cdot 2\left( \sideset{_\langle}{}{\mathop D} \sigma_{\mu \nu \rangle} + \frac1{\tilde d - 1} \sigma_{\mu\nu} \partial u \right) + \tilde \lambda_1 \cdot 4\sigma_{\langle \mu}^{~~\rho}\sigma_{\nu \rangle \rho} + \tilde \lambda_2 \cdot 2\sigma_{\langle \mu}^{~~\rho}\Omega_{\nu \rangle \rho},
\end{align}
where we also write the 2nd order coefficients of AdS$_{\tilde d + 1}$ black hole background with a tilde. Also note that here we do not keep quantities with dimension like $r_H$ and $L$, since the Kanitscheider-Skenderis proposal only involves with numerical value of the 2nd order transport coefficients. After using the transformation rule (\ref{eq: Kanitscheider-Skenderis transformation}), the above will become
\begin{align}
  & \widetilde{\eta\tau_\pi} \cdot 2\left( \sideset{_\langle}{}{\mathop D} \sigma_{\mu \nu \rangle} + \frac1{d-1} \sigma_{\mu\nu} \partial u \right) - (d-1) \chi \widetilde{\eta\tau_\pi} \frac{2 \sigma_{\mu\nu} \partial u}{d-1} + \tilde \lambda_1 \cdot 4\sigma_{\langle \mu}^{~~\rho}\sigma_{\nu \rangle \rho} \cr
  & + \tilde \lambda_2 \cdot 2\sigma_{\langle \mu}^{~~\rho}\Omega_{\nu \rangle \rho} + P_{\mu\nu} \left( 2\chi \widetilde{\eta\tau_\pi} D(\partial u) + \frac{2 \chi \widetilde{\eta\tau_\pi}}{\tilde d -1} (\partial u)^2 + \chi \tilde \lambda_1 \cdot 4\sigma_{\alpha \beta}^2 \right).
\end{align}
Thus we get the relations between the 2nd order transport coefficients of nonconformal fluid and that of the conformal fluid:
\begin{align}\label{eq: tranport coefficients between AdS(d+1) to Dp}
  \eta\tau_\pi &= \widetilde{\eta\tau_\pi}, \quad \eta\tau_\pi^* = - (d-1) \chi \, \widetilde{\eta\tau_\pi}, \quad \lambda_1 = \tilde \lambda_1, \quad \lambda_2 = \tilde \lambda_2, \cr
  \zeta\tau_\Pi &= 2 \chi \, \widetilde{\eta\tau_\pi}, \quad \xi_1 = \chi \, \tilde \lambda_1, \quad \xi_2 = \frac{2\chi}{\tilde d -1} \widetilde{\eta\tau_\pi}.
\end{align}

Now we will use the results of D1, D2 and D4-brane to test whether the relations (\ref{eq: tranport coefficients between AdS(d+1) to Dp}) are reliable. Since the D4-brane result can be got from AdS$_7$ black hole background, so we have $\tilde d =6$, $d=p+1=5$ and $\chi=1/20$. The 2nd order transport coefficients for AdS$_7$ black hole are
\begin{align}
  \widetilde{\eta\tau_\pi} = \frac12 + \frac16 H_{\frac13}, \quad \tilde \lambda_1 = \frac12, \quad \tilde \lambda_2 = -1 + \frac13 H_{\frac13}.
\end{align}
Thus we can get from (\ref{eq: tranport coefficients between AdS(d+1) to Dp}) that
\begin{align}
  \eta\tau_\pi &= \frac12 + \frac16 H_{\frac13}, \quad \eta\tau_\pi^* = - \frac{1}{10} - \frac{1}{30} H_{\frac13}, \quad \lambda_1 = \frac12, \quad \lambda_2 = -1 + \frac13 H_{\frac13}, \cr
  \zeta\tau_\Pi &= \frac{1}{20} + \frac{1}{60} H_{\frac13}, \quad \xi_1 = \frac{1}{40}, \quad \xi_2 = \frac{1}{100} + \frac{1}{300} H_{\frac13}.
\end{align}
Since a global scaling factor for the above results do not matter, we can multiply a scaling factor 2 to the above results and get
\begin{align}
  \eta\tau_\pi &= 1 + \frac13 H_{\frac13}, \quad \eta\tau_\pi^* = - \frac15 - \frac{1}{15} H_{\frac13}, \quad \lambda_1 = 1, \quad \lambda_2 = -2 + \frac23 H_{\frac13}, \cr
  \zeta\tau_\Pi &= \frac{1}{10} + \frac{1}{30} H_{\frac13}, \quad \xi_1 = \frac{1}{20}, \quad \xi_2 = \frac{1}{50} + \frac{1}{150} H_{\frac13}.
\end{align}
Compare the above with (\ref{eq: Dp 2nd order coefficients}) for the case $p=4$, we can see that $\eta\tau_\pi,\zeta\tau_\Pi,\lambda_{1,2},\xi_1$ are the same as the results of D4-brane, but $\eta\tau_\pi^*$ and $\xi_2$ do not match.

By the same token, one can check the situations of D1 and D2-brane. The reduced bulk gravity of D1-brane relates to AdS$_{4}$ black hole background, after using the relations (\ref{eq: tranport coefficients between AdS(d+1) to Dp}) and multiplying a scaling factor 2 one can reproduce exactly the second order transport coefficients for D1-brane. The second order transport coefficients of D2-brane should be got from that of AdS$_{6}$ black hole by using (\ref{eq: tranport coefficients between AdS(d+1) to Dp})\footnote{One will need the reflection relation of Harmonic number $H_{1-\alpha}=H_\alpha + \pi\cot\pi\alpha + \frac{1}{1-\alpha} - \frac1\alpha$ in comparing the results got from (\ref{eq: tranport coefficients between AdS(d+1) to Dp}) with the second order transport coefficients of D2-brane in (\ref{eq: Dp 2nd order coefficients}).}. But one will see that the results got this way can not match any transport coefficient of D2-brane in (\ref{eq: Dp 2nd order coefficients}).

The authors of \cite{Kleinert1610} derive out the exact analytical results of the second order transport coefficients for 5d Chamblin-Reall background and use the leading nonconformal corrections of the results to compare with \cite{Bigazzi1006}. It is interesting to note that they also find $\eta\tau_\pi^*$ and $\xi_2$ in \cite{Bigazzi1006} are wrong. The reason they think is that \cite{Bigazzi1006} uses (\ref{eq: wrong relations in Romatschke0906}) (first proposed in \cite{Romatschke0906}) to get $\eta\tau_\pi^*$ and $\xi_2$ whose correct form should be (\ref{eq: Kleinert-Probst relation}). Considering that \cite{Romatschke0906} derived the relations (\ref{eq: wrong relations in Romatschke0906}) also using the method of \cite{Kanitscheider0901}, we may draw the conclusion that the method of getting the 2nd order transport coefficients of nonconformal Dp-brane from AdS black hole background as suggested in \cite{Kanitscheider0901} is applicable only for some special cases. It is better to use the standard methods like the Minkowskian AdS/CFT prescription or the fluid/gravity correspondence to calculate the 2nd order transport coefficients for nonconformal backgrounds.

\subsection{Dispersion relations}\label{subsec: dispersion relation}

At last, we would like to offer the dispersion relations for the dual relativistic fluid of Dp-brane with $1\leq p\leq 4$. One needs to expand $r_H(x)$ and $\beta_i(x)$ as $r_H(x) = r_H(0) + \delta r_H \mathrm{e}^{ik_\mu x^\mu}$ and $\beta_i(x) = \delta\beta_i \mathrm{e}^{ik_\mu x^\mu}$, respectively. Then put them into the conservation equation of the stress-energy tensor of Dp-brane $\partial^\mu T_{\mu\nu} = 0$ and find the coefficient matrix of the vector $(\delta r_H,\delta \beta_i)^T$ where the subscript ``T" stands for taking the transpose. Then the determinant of that coefficient matrix equate to 0 will give two equations from which we can solve the transverse and the longitudinal modes of the dispersion relations (denoted separately as $\omega_T$ and $\omega_L$). The result is
\begin{align}
  \omega_T =& -i \frac{1}{7-p} \frac{L_p^\frac{7-p}2}{r_H^\frac{5-p}2} k^2 - i \frac{(7-p)+(5-p) H_\frac{5-p}{7-p}}{(5-p)(7-p)^3} \frac{L_p^\frac{3(7-p)}2}{r_H^\frac{3(5-p)}{2}} k^4, \quad (p\neq1) \\
  \omega_L =& \pm \sqrt{\frac{5-p}{9-p}} k - i \frac{4}{(7-p)(9-p)} \frac{L_p^\frac{7-p}{2}}{r_H^\frac{5-p}{2}} k^2
  \pm \frac{4 (5-p)^\frac12 \Big(1+ H_\frac{5-p}{7-p} \Big)}{(7-p)^2 (9-p)^\frac32} \frac{L_p^{7-p}}{r_H^{5-p}} k^3 \cr
  & - i \frac{32 \Big(7-p + (5-p) H_\frac{5-p}{7-p} \Big)}{(5-p) (7-p)^3 (9-p)^2} \frac{L_p^\frac{3(7-p)}{2}}{r_H^\frac{3(5-p)}{2}} k^4.
\end{align}
Here $k=|\vec k|$ and also note that $p$ can not be 1 for the transverse mode.

\section{Summary and outlook}\label{sec: summary}

In this paper, we have derived all the dynamical second order transport coefficients for Dp-brane using fluid/gravity correspondence. D5 and D6-brane do not have physical dual relativistic fluid; D3-brane is dual to 4d conformal fluid of which the results are already well known in the literature; thus the new results are mainly for D1, D2 and D4-brane, which correspond to nonconformal fluid of dimension 2, 3 and 5.

The validity of the identities among 2nd order transport coefficients has exceptions: $\tau_\pi = \tau_\Pi$ is not valid for D1-brane; the HY relation, the generalized Romatschke (\ref{eq: relation between xi_1 and lambda_1 (generalized)}) and KP (\ref{eq: Kleinert-Probst relation (generalized)}) relations are appropriate for Dp-brane with $p>2$, or for relativistic fluid of dimension $d>3$. The general form of the 2nd order transport coefficients of Dp-brane (\ref{eq: Dp 2nd order coefficients}) satisfies the HY, Romatschke and KP relations. We also think that the constraint relations between the 2nd order transport coefficients, which were first proposed in \cite{Banerjee1203} and rewritten in a more useful form in \cite{Moore1210}, need to be generalized if the relativistic fluid is not in 4d. So one should be careful when using them in dimensions other than 4.

This work leaves us some interesting problems to explore. First, we can continue to study the thermodynamical 2nd order transport coefficients for Dp-brane via the real-time AdS/CFT correspondence. Second, D5 and D6-brane are not dual to any relativistic fluid. So we may need to use the NS5-brane if we want to study the relativistic fluid of dimension larger than 5. NS5-brane, if it has, will be dual to a nonconformal relativistic fluid of 6d. Finally, we can tell from the results of D4-brane and compactified D4-brane that the compactification on the brane will change the spatial dimension of the fluid thus the 2nd order transport coefficients relating viscous scalars, \textit{i.e.}, $\eta\tau_\pi^*$, $\zeta\tau_\Pi$ and $\xi_{1,2}$. So we can compactify more directions on D4-brane to get nonconformal fluids of dimension 2 and 3, and one can also compactify the other Dp-branes. It will be interesting to see whether D3-brane will correspond to a nonconformal fluid after compactification.

Through \cite{Wu1508,Wu1604} and this work, we have successfully generalized the original discussion of fluid/gravity correspondence to nonconformal regime for various dimensions. By the stress-energy tensor of general dimension (\ref{eq: Dp 2nd order stress-energy tensor (general)}), we actually offers a group of nonconformal relativistic fluid models. We hope these models can be used to get the analytic solutions by using the framework of \cite{Gubser1006,Gubser1012,Marrochio1307,Hatta1401,Hatta1403,Nagy0709,Jiang1711} thus will give us more hints on the phenomena of relativistic heavy ion collider.

\section*{Acknowledgement}

First, we would like to thank Yu Lu for his generous help on Mathematica code and programming. We also want to thank Zoltan Bajnok, Haryanto Siahaan for very helpful discussions, Ze-Fang Jiang for introducing some references on exact solutions in relativistic hydrodynamics, and Michael Abbott for his kind help on improving presentation. This work is supported by the Hungarian National Science Fund NKFIH (under K116505), a Lend\"ulet Grant and the Young Scientists Fund of the National Natural Science Foundation of China (Grant No. 11805002).

\appendix

\section{Dimensional reduction of bulk metric}\label{app: reduction of bulk metric}

We will do the dimensional reduction on the following 10d metric
\begin{align}
  ds^2 = e^{2\alpha_1 A} g_{MN} dx^M dx^N + L_p^2 e^{2\alpha_2 A} d\Omega_{8-p}^2,
\end{align}
where the dimension of the $\{x^M\}$ part is $(p+2)$-dimensional, $A=A(x^M)$ only depend on coordinates of the first $p+2$ dimensions, the second part is a $(8-p)$-dimensional unit sphere with the metric $\gamma_{ab}$. $\alpha_{1,2},~L_p$ are all constants. The components of connection are
\begin{align}
  \widehat \Gamma^M_{NP} &= \Gamma^M_{NP} + \alpha_1(\delta^M_N \partial_P A + \delta^M_P \partial_N A - g_{NP}\nabla^M A), \\
  \widehat \Gamma^M_{ab} &= -\alpha_2 L_p^2 e^{(-2\alpha_1+2\alpha_2)A} \nabla^M A \gamma_{ab}, \\
  \widehat \Gamma^a_{M b} &= \alpha_2 \partial_M A \delta^a_b, \\
  \widehat \Gamma^a_{bc} &= \Gamma^a_{bc}.
\end{align}
The Ricci tensor is
\begin{align}
     \widehat R_{MN} &= R_{MN} - [p\alpha_1 + (8-p)\alpha_2] \nabla_M\nabla_N A - \alpha_1 g_{MN} \nabla^2 A + \big[ p\alpha_1^2 + 2(8-p) \alpha_1\alpha_2 \cr
   & - (8-p)\alpha_2^2 \big] \nabla_M A \nabla_N A - \left[ p\alpha_1^2 + (8-p)\alpha_1\alpha_2 \right] g_{MN} (\nabla A)^2, \\
   \widehat R_{ab} &= (7-p) \gamma_{ab} - L_p^2 e^{(-2\alpha_1+2\alpha_2)A} \big[ \alpha_2 \nabla^2 A + \left( p \alpha_1\alpha_2 + (8-p)\alpha_2^2 \right) (\nabla A)^2 \big] \gamma_{ab}.
\end{align}
The Ricci scalar is
\begin{align}
  \widehat R &= \frac{(7-p)(8-p)}{L_p^2} e^{-2\alpha_2 A} + e^{-2\alpha_1 A} \left[ R - \left( (2p+2)\alpha_1 + (16-2p)\alpha_2 \right) \nabla^2 A \right.\cr
  &\left. - \left( (p^2+p)\alpha_1^2 + (16p-2p^2)\alpha_1\alpha_2 + (8-p)(9-p)\alpha_2^2 \right) (\nabla A)^2 \right]
\end{align}
The quantities with a hat like $\widehat \Gamma$ and $\widehat R$ are 10-dimensional quantities while those without like $R$ are $(p+2)$-dimensional ones.


\providecommand{\href}[2]{#2}\begingroup\raggedright\endgroup

\end{document}